\documentclass[aps, twocolumn, letterpaper, superscriptaddress, balancelastpage, longbibliography, preprintnumbers]{revtex4-2}

\usepackage{graphicx}% Include figure files
\usepackage{dcolumn}% Align table columns on decimal point
\usepackage{amsmath, amsthm, mathrsfs, amsfonts, dsfont}
\usepackage{booktabs}
\usepackage{array}
\usepackage{url}
\usepackage{xcolor}
\usepackage[colorlinks,citecolor=blue,urlcolor=blue,linkcolor=blue]{hyperref}
\usepackage{bm, bbold, xcolor, slashed}
\usepackage{easy-todo}
\usepackage{stmaryrd}
\usepackage{braket}
\usepackage{graphicx}
\usepackage{amsmath,amssymb}
\usepackage{amsfonts}
\usepackage{mathtools}
\usepackage{tcolorbox}
\usepackage{color}
\usepackage{xcolor}
\usepackage{tikz}
\usepackage{cases}
\usepackage{url}
\usepackage{bm}
\usepackage{here}
\usepackage{physics}
\usepackage{mathrsfs}
\usepackage{comment}
\usepackage{empheq}
\usepackage{cancel}
\usepackage[compat=1.1.0]{tikz-feynhand}
\usetikzlibrary{decorations.pathreplacing, tikzmark}

\usepackage{ytableau}
\ytableausetup{mathmode, boxsize=1.5em, aligntableaux = center}

\newcommand{\bub}[5]{
\filldraw [fill=white] (#1,#2) circle [radius=#3];
\node at (#1,#2) {#5 $#4$};
}

\newcommand{\Amp}{\mathcal{A}}
\newcommand{\nn}{\nonumber \\}
\newcommand{\hd}{\hat{\rm d}}
\newcommand{\hdelta}{\hat{\delta}}
\newcommand{\Mpl}{M_{\rm Pl}}
\newcommand{\Disc}{{\rm Disc}}
\newcommand{\Ycal}{\mathcal{Y}}

\begin{document}

\title{
Black Hole Thermodynamics Meets On-Shell Amplitudes: \\
\scalebox{0.9}{Local Detailed Balance and Thermal Spectrum from Spin Universality and Unitarity}
}

\author{Dogan Akpinar}
\email{dogan.akpinar@ed.ac.uk} 
\affiliation{Centre for Mathematical Sciences, Plymouth University, Plymouth, PL4 8AA, UK}
\affiliation{Higgs Centre for Theoretical Physics, School of Physics and
Astronomy, University of Edinburgh, Edinburgh, EH9 3FD, UK} 
\author{Katsuki Aoki}
\email{katsukiaoki@mail.saitama-u.ac.jp} 
\affiliation{Graduate School of Science and Engineering, Saitama University, 255 Shimo-Okubo, Sakura-ku, Saitama 338-8570, Japan}
\affiliation{Center for Gravitational Physics and Quantum Information, Yukawa Institute for Theoretical Physics, Kyoto University, 606-8502, Kyoto, Japan}
\author{Andrea Cristofoli}
\email{cristofoli@yukawa.kyoto-u.ac.jp} 
\affiliation{Center for Gravitational Physics and Quantum Information, Yukawa Institute for Theoretical Physics, Kyoto University, 606-8502, Kyoto, Japan}
\author{Hyun Jeong}
\email{jeong\_hyun@resceu.s.u-tokyo.ac.jp} 
\affiliation{Department of Physics, Graduate School of Science, The University of Tokyo, Tokyo 113-0033}
\affiliation{Research Center for the Early Universe (RESCEU), Graduate School of Science, The University
of Tokyo, Tokyo 113-0033, Japan}
\affiliation{Kavli IPMU (WPI), UTIAS, The University of Tokyo, Kashiwa, Chiba 277-8583, Japan}
\author{Kaho Yoshimura}
\email{yoshimura-kaho848@g.ecc.u-tokyo.ac.jp} 
\affiliation{Graduate School of Arts and Sciences, University of Tokyo, Komaba, Meguro-ku, Tokyo 153-8902}

\date{\today}

\begin{abstract}
    We develop an on-shell framework for thermal dissipation and radiation by macroscopic objects, whose large degeneracy of internal states is encoded in their entropy. In this framework, equilibrium asymptotic states are represented as on-shell particles, while non-equilibrium processes are described by on-shell transition amplitudes between them. A central observation is that spinning states remain essential even for macroscopically non-rotating objects. Consistency with macroscopic symmetries then implies spin universality, whereby all spinning states are governed by a single universal coupling. A key consequence is that absorption and emission probabilities are controlled by the same coupling, yielding local detailed balance directly from on-shell data. Applied to black holes, our framework reproduces the thermal emission spectrum and relates the Hawking temperature to the condition of maximal absorption consistent with unitary time evolution.
\end{abstract}

\preprint{STUPP-26-299}
\preprint{YITP-25-143}
\preprint{RESCEU-18/26}
\preprint{IPMU26-0025}

\maketitle

%----------------------------------------------------------------------------

\textbf{\textit{Introduction---}}Black holes (BHs) are conventionally understood as classical solutions of the Einstein field equations. At the quantum level, however, this description gives rise to profound questions concerning unitarity; a notable example being the information loss paradox~\cite{PhysRevD.14.2460}. In light of this inconsistency, a more radical viewpoint---dating back to 't Hooft~\cite{tHooft:1984kcu, tHooft:1996rdg}---is to instead regard BHs as quantum-mechanical states, whose dynamics are described by a unitary $S$-matrix.

This approach has recently gained renewed attention, driven by applying modern amplitude techniques to gravitational wave physics; see, e.g., Refs.~\cite{Bern:2019nnu, Bern:2025wyd, Driesse:2024feo, Driesse:2024xad, Driesse:2026qiz, Dlapa:2026oyq, Akpinar:2024meg, Akpinar:2025bkt, Haddad:2025cmw, Bohnenblust:2024hkw, Aoude:2023vdk,Akpinar:2025byi,Bautista:2026qse, Bjerrum-Bohr:2026fhs, Bjerrum-Bohr:2026fhx}. As of late, this program has been extended to encompass the quantum aspects of BHs~\cite{Aoude:2024sve,Ilderton:2025umd,Aoki:2025ihc,Ilderton:2025aql,Aoude:2025jvt,Carrasco:2025ymt,Carrasco:2025bgu,Aoude:2026bek, Carrasco:2026ijt}, thereby leading to a refined formulation of the BH $S$-matrix. Nevertheless, these approaches typically assume semiclassical physics from the outset in order to invoke the existence and properties of a BH horizon. While horizon absorption and emission can be described using on-shell amplitudes---allowing BHs to be treated as particles~\cite{Aoude:2023fdm,Jones:2023ugm, Chen:2023qzo, Aoude:2024jxd, Bautista:2024emt, Aoki:2025ihc, Gatica:2025uhx}---this naturally raises the question of what distinguishes BHs from other compact objects or microscopic particles. More broadly, it remains unclear whether BH physics is tied more deeply to fundamental properties of scattering amplitudes, such as unitarity itself.

In this \textit{Letter}, our overarching goal is to understand the origin of macroscopic thermal and dissipative phenomena of BHs, or more generically of macroscopic objects, from the general properties of on-shell amplitudes. More concretely, we model a macroscopic object in thermal equilibrium as a particle endowed with a large degeneracy of internal states. Absorption or emission of external particles is then described as a scattering process in which the object transitions between distinct equilibrium states. From the contemporary amplitudes perspective, all essential physical information is encoded solely in the asymptotic in- and out-states. Explicit intermediate time evolution and bulk non-equilibrium dynamics therefore play no fundamental role. In this sense, the central feature of our approach is that dissipative and non-equilibrium phenomena are reformulated as transitions between on-shell states.

A central observation is that spinning states remain indispensable even when describing macroscopically non-rotating objects, such as Schwarzschild BHs. As a first step, we systematically classify unequal-mass three-point amplitudes and discuss the assumptions underlying our treatment of microstates. Requiring consistency of absorption and emission processes with macroscopic spherical symmetry leads to an explicit emergence of spin universality, whereby all spinning equilibrium states are governed by a single universal coupling. This universality, in turn, implies that absorption and emission probabilities are controlled by the same on-shell data, namely a single universal coupling, thereby yielding local detailed balance---an important ingredient of non-equilibrium statistical physics~\cite{10.21468/SciPostPhysLectNotes.32}---directly from the $S$-matrix. These general considerations are sufficient to recover the thermal spectrum, which emerges from the interplay between macroscopic scattering data and the spin-universal structure of microscopic amplitudes; this makes no reference to semiclassical bulk evolution. Finally, our approach rediscovers a profound connection between horizon physics and thermal behavior through a saturation of unitarity.

In the remainder of this \textit{Letter}, we develop the formalism underlying these results. Alongside this, the appendices contain a summary of our conventions, a relation to the coarse-grained description, a detailed derivation of absorption probabilities and spin universality, and a correspondence between classical transmissivities and quantum transition probabilities.

\textbf{\textit{Isolated objects as particles---}}Let us assume that, akin to quantum states like electrons and pions, macroscopic objects can be represented by on-shell one-particle states $\ket{p,s,v}$, labeled by its momentum $p$ with $p^2=M^2$, its spin quantum number $s= n/2$ for positive-integer $n$, and an internal label $v$. This extra label $v$ characterizes additional degrees of freedom that distinguishes microscopic states sharing the same macroscopic data, such as mass and spin. Spinning states may equivalently be written as $\ket{p,{\{I_{2s}\}},v}$, where $\{I_{2s}\}=\{I_1\cdots I_{2s}\}$ denotes the symmetrized little-group indices. 

In the context of statistical mechanics, the microstates of a microcanonical ensemble are treated as equally probable.  Following this coarse-grained description, we may treat microstates collectively, in which the internal label need not be resolved~\footnote{Scattering amplitudes are agnostic to theory-dependent descriptions. More precisely, information about the microscopic theory is encoded in exclusive probabilities, while the macroscopic details are in inclusive probabilities, the latter of which corresponds to the notion of coarse-graining. See Appendix~\ref{sec:density_matrix} for more details on this point.}. Then, we denote the resulting macroscopic one-particle state by $\ket{p,s}$, while the degeneracy of unresolved internal states is encoded in some spectral density $\rho(\mu,s)$, where $\mu$ denotes the mass parameter integrated over. The completeness relation for macroscopic objects then takes the form \footnote{This type of completeness relation is used to model BHs and successfully describes horizon absorption, Hawking radiation, and even BH formations \cite{Giddings:2007qq,Giddings:2009gj,Aoude:2023fdm,Jones:2023ugm, Chen:2023qzo, Aoude:2024jxd, Bautista:2024emt, Aoki:2024boe, Aoki:2025ihc, Gatica:2025uhx}.}
\begin{align}
\label{eq:spec-dens}
    \hat{1}=\sum_{s}\int \dd \mu^2\, \int \dd\Phi(p) \rho(\mu,s) \ket{p,s}\bra{p,s}
    \,,
\end{align}
with $\dd \Phi(p)$ denoting the on-shell integral. We consider a continuous spectrum $\rho=C(\mu,s)e^{\mathcal{S}(\mu,s)}$ where the exponential growth $\mathcal{S}(\mu,s)$ is identified with the microcanonical entropy, while $C(\mu,s)$, assumed to vary at most polynomially in $\mu$ and $s$, depends on which mass interval we count states to define the entropy. 

The spinning dynamics of a macroscopic object are often described by a classical spin vector $S^\mu$, carrying both a magnitude and direction. Within the scattering amplitudes description, this quantity emerges through an expectation value of the Pauli--Lubanski pseudovector~\cite{Maybee:2019jus}. Although several equivalent descriptions of spin exist in the amplitudes literature, it will be convenient to employ coherent spin states~\cite{Aoude:2021oqj}
\begin{align}
\ket{p, \alpha}:=e^{-\frac{1}{2}\|\alpha\|^2} \sum_{2s=0}^{\infty} \frac{\alpha^{\{I_{2s}\}}}{\sqrt{(2s)!}} \ket{p,\{I_{2s} \}}
\,,
\label{coherent_spin}
\end{align}
where $\tilde{\alpha}_I=(\alpha^I)^*, \| \alpha \|^2 = \tilde{\alpha}_I \alpha^I$ and $\alpha^{\{ I_{2s} \}} =\alpha^{I_1} \cdots \alpha^{I_{2s}}$. The classical spin vector is computed as $S^{\mu}=\mathcal{O}\left(\hbar \| \alpha \|^2\right)$
where we temporarily restore $\hbar$ to make the classical limit explicit.

In this \textit{Letter}, we focus on objects without macroscopic rotation, such as Schwarzschild BHs, corresponding to spins that satisfy $|S|\ll GM^2$, with $|S|^2 = -S^2$~\footnote{This corresponds to neglecting the spin in the general-relativistic sense, which we take for an illustrative purpose.}. The spinless dynamics then arises from the leading contribution, i.e.~$\mathcal{O}(S^0)$, in a small-classical-spin expansion. Importantly, however, we define non-rotating macroscopic objects as the small-classical-spin limit of spinning states, rather than as strictly spin-$0$ particles. There are two main reasons for this choice. First, even in the absence of macroscopic rotation, a macroscopic object is expected to carry non-negligible quantum spin, i.e.,~$|S|\sim\hbar$. 
Second, more importantly, our framework treats initial and final states on equal footing. Macroscopically, a Schwarzschild BH remains approximately Schwarzschild after absorbing a partial wave of angular momentum $\ell$. By contrast, a spin-$0$ particle would transition into a spin-$\ell$ state after the absorption, placing the initial and final states in different representations. For these reasons, we will consider objects with $1 \ll |\alpha| \ll M/\Mpl$ and focus on their leading $\mathcal{O}(S^0)$ dynamics, for which the dominant contributions arise from states satisfying $1 \ll s \ll M^2/\Mpl^2$. We henceforth refer to such objects as \textit{slowly rotating objects}. As we will see, the use of spinning states is in fact essential for recovering macroscopic physics, even in the slow-rotation limit.

It is crucial to note that our discussion solely considers scattering problems in infinite volumes. Strictly speaking, these objects only remain in equilibrium when exactly free; once coupled to photons, they lose their mass through thermal radiation. In fact, this is closely analogous to the situation for unstable particles: they are absent in asymptotic states by unitarity~\cite{Veltman:1963th, Denner:2014zga}. However, for processes whose characteristic timescales are much shorter than the lifetime of the object, perturbation theory allows them to be treated effectively as asymptotic states. This assumption forms a central ingredient of the present \textit{Letter}.

\textbf{\textit{Spinning three-point amplitudes---}}The dynamics of our macroscopic objects will be described by on-shell amplitudes. Specifically, we consider processes by which an object either absorbs or emits a particle and changes its macroscopic parameters. For simplicity, we will only consider absorption or emission of massless bosons, such as a scalar, a photon, or a graviton, described by three-point amplitudes that involve unequal masses and spins: $\Amp_{\{ I_{2s_1}\}}{}^{\{J_{2s_2}\}}(p_1;p_2;k^h)$. Here $I$ and $J$ are the little group indices of the massive particles 1 and 2, respectively, and $k$ is the momentum of the absorbed/emitted particle of helicity $h$. We adopt the all-ingoing notation when writing the amplitudes, meaning that $p^0>0$ represents an ingoing particle while $p^0<0$ is outgoing.

There are $s_1+s_2-|s_1-s_2|+1=\min(2s_1+1,2s_2+1)$ independent three-point amplitudes~\cite{Arkani-Hamed:2017jhn}. We first classify these amplitudes in terms of kinematics. We assume $M_2>M_1$ and introduce
\begin{align}
    \xi^I_-&= \frac{\langle kp_1^I \rangle}{\sqrt{M_2^2-M_1^2}}\,, \qquad
    \xi_+^I= -\frac{ [ kp_1^I ]}{\sqrt{M_2^2-M_1^2}}, \\
    \xi_-^J&= \frac{\langle kp_2^J \rangle}{\sqrt{M_2^2-M_1^2}}\,, \qquad
    \xi_+^J= \frac{ [ kp_2^J ]}{\sqrt{M_2^2-M_1^2}}\,,
\end{align}
and 
\begin{align}
    \mathcal{E}^{IJ}&=\xi_+^I \xi_-^J-\xi_+^J\xi_-^I\,, \label{Ecal} \\
    {}_{h} \Ycal^{A_1 \cdots A_{2\ell}}&=\mathcal{N}_{\ell,h}\, \xi_-^{\{ A_1} \cdots \xi_-^{A_{\ell -h}}  \xi_+^{A_{\ell-h+1}} \cdots \xi_+^{A_{2\ell}\}}
    \,, \label{Ycal}
\end{align}
where $\mathcal{N}_{\ell,h}=\sqrt{\frac{2\ell+1}{4\pi}\frac{(2\ell)!}{(\ell!)^2(\ell+h)!(\ell-h)!}} $
and $A$ denotes either type of little-group index, i.e.~$I$ or $J$. One can show that $\mathcal{E}^{IJ}$ and ${}_{h} \Ycal^{A_1 \cdots A_{2\ell}}$ agree with the Levi-Civita symbol and the spin-weighted spherical harmonics ${}_hY_{\ell, m}$ in the three-point kinematic with the normalization
\begin{align}
    {}_{h'} \Ycal_{A_1 \cdots A_{2\ell}}(\hat{\bm{k}}')\, {}_{h} \Ycal^{A_1\cdots A_{2\ell}}(\hat{\bm{k}}) & = (-1)^{\ell+h}\frac{2\ell +1}{4\pi} \mathcal{P}_{h',h}^{(\ell)} \,,
\end{align}
where $\mathcal{P}_{h',h}^{(\ell)}$---which we will refer to as spinning Legendre polynomials---coincide with the usual Wigner-d function up to little-group freedom. Given \eqref{Ecal} and \eqref{Ycal}, we identify each independent structure of generic massive-massive-massless three-point amplitudes with ${}_hY_{\ell, m}$ of $|s_1-s_2|\leq \ell \leq s_1+s_2$~\footnote{We use spherical harmonics as the slow-rotation limit isolates the spin-monopole dynamics, which describes non-rotating classical objects. To describe rotating classical objects, one must instead employ spin-weighted spheroidal harmonics.}:
\begin{align}
\Amp_{\{ I_{2s_1}\}}{}^{\{J_{2s_2}\}} = \sum_{\ell = |s_1-s_2|}^{s_1+s_2} \Amp^{(\ell)}_{\{I_{2s_1}\}}{}^{\{J_{2s_2}\}}\label{eq:3-point-important}\,,
\end{align}
with 
\begin{align}
    \Amp^{(\ell)}_{\{I_{2s_1}\}}{}^{\{J_{2s_2}\}} & =  g^{(\ell)}_{s_1,s_2,h}\,\Ycal_{\{ I_{1} \cdots I_{\ell+s_1-s_2}}{}^{\{ J_{1} \cdots J_{\ell+s_2-s_1}} \nonumber \\
& \hspace{-0.1cm}\times \mathcal{E}_{ I_{\ell+s_1-s_2+1}}{}^{J_{\ell+s_2-s_1+1}} \cdots \mathcal{E}_{ I_{2s_1}\} }{}^{J_{2s_2}\}} \,.
\label{3pt_ell}
\end{align}
Here \eqref{3pt_ell} is the unique amplitude that describes a transition between a spin-$s_1$ state and a spin-$s_2$ state through the absorption $(\omega>0)$ or the emission $(\omega<0)$ of a spherical wave with angular momentum $\ell$. Note that \eqref{3pt_ell} represents a transition amplitude for specific microstates, which we will call an exclusive process. To discuss this, we temporarily introduce the microstate labels $v$ and denote the coupling as $g_{v_1,v_2}$ while suppressing other labels. Since we have considered degenerate microstates, the exclusive probabilities for different microstates are assumed to be the same, which results in the couplings having the same modulus squared $|g_{v_1,v_2}|^2=|g_{v_1',v_2'}|^2$. Importantly, this does not necessarily imply that they share the same phase, i.e.,~$g_{v_1,v_2}= e^{i\theta_{v_1,v_2}}g$ with $g$ being the common modulus. See Appendix~\ref{sec:density_matrix} for more explanations, with a brief comparison of our assumptions with the eigenstate thermalization hypothesis~\cite{Deutsch:1991msp,Srednicki:1994mfb,Deutsch:2018ulr,Srednicki:1999bhx,Rigol:2012yjf}.

To reduce clutter, we omit the microstate label, while keeping in mind that the couplings may carry different phases.

\textbf{\textit{Spin universality and local detailed balance---}}Having established the relevant amplitudes, we now turn to our central results: the emergence of spin universality and local detailed balance. These results follow from the requirement that microscopic spinning states are consistent with macroscopic symmetries.

Our focus is on inclusive probabilities---total probabilities for transitions to all possible microstates---associated with the absorption or emission of a single quanta of angular momentum $\ell$. By unitarity, these probabilities are encoded in the partial-wave discontinuity of the 2-to-2 amplitude. In the case of absorption,
\begin{align}
    ~{\rm Disc} \frac{\mathcal{A}_{1'\leftarrow 1}^{-h,h}}{2\pi i }&= e^{-\| \alpha\|^2} \sum_{s_1,s_1',s_2} 
 \frac{\rho(M_2,s_2)}{\sqrt{(2s_1)!(2s_1')!}} \\
&\times  \tilde{\alpha}_{\{ I'_{2s'_1} \}}\tilde{\Amp}^{\{I'_{2s'_1}\}}{}_{\{ J_{2s_2}\}}
 \Amp_{\{I_{2s_1}\}}{}^{\{J_{2s_2}\}}
\, \alpha^{\{ I_{2s_1} \}} \nonumber \,,
\end{align}
where $\Amp$ and $\tilde{\Amp}$, with $\tilde{\Amp}$ being the conjugate amplitude, are functions of $(p_1', k')$ and $(p_1, k)$, and $1\leftrightarrow 2$ with $h\leftrightarrow -h$ for emission. For concreteness, we will consider the absorption process in what follows. When absorbing a partial wave $\ell$, the definite spin state can change by discrete values of $\Delta s=s_2-s_1=0,\pm 1, \cdots \pm \ell$. In such a process, initial states of different spins can transition into the same state, e.g.,~$s_1=s \to s_2=s$ and/or $s_1=s-\Delta s\to s_2=s$, which results in interference terms that are, in general, proportional to $g^*_{s,s}g_{s-\Delta s,s}$. For a highly degenerate final state, $\rho(M_2,s_2)\gg 1$, these interference terms are suppressed relative to the diagonal contributions, provided that the microscopic couplings have uncorrelated phases across the degenerate microstates; they can survive only if the phases are coherently aligned. We assume they are not in phase and thus obtain
\begin{align}
    ~\frac{{\rm Disc} \mathcal{A}_{1'\leftarrow 1}^{-h,h}}{2\pi i }&= e^{-\| \alpha\|^2} \sum_{s_1,s_2,\ell} 
 \frac{\rho(M_2,s_2)}{(2s_1)!} ||\mathcal{A}^{(\ell)}_{s_1,s_2,h}||^2\,,
\label{Disc_abs}
\end{align}
with
\begin{equation}
    \label{eq:AModulusSquared}
    ||\mathcal{A}^{(\ell)}_{s_1,s_2,h}||^2 = 
\tilde{\alpha}_{\{ I'_{2s_1} \}} \tilde{\Amp}_{(\ell)}^{\{I'_{2s_1}\}}{}_{\{J_{2s_2}\}} \Amp^{(\ell)}_{\{I_{2s_1}\}}{}^{\{J_{2s_2}\}}\, \alpha^{\{ I_{2s_1} \}}\,.
\end{equation}
Evaluating the discontinuity then boils down to computing \eqref{eq:AModulusSquared}. Following the notation and discussion of Appendix~\ref{sec:probability}, 
in the limit of a slow-rotating object, \eqref{eq:AModulusSquared} is given by
\begin{align}
     ||\mathcal{A}_{s_1, s_2,h}^{(\ell)}||^2& \simeq (-1)^{\ell -\Delta s} |g_{s_1, s_1+\Delta s, h}^{(\ell)}|^2 ||\alpha||^{4s_1 - 4\ell} \\
     & \hspace{0.5cm} \times {}_{-h}\tilde{\Ycal}^{\tilde{\alpha}^{\,\ell -\Delta s}}{}_{\alpha^{\,\ell + \Delta s}} \ {}_{h}\Ycal_{\alpha^{\,\ell - \Delta s}}{}^{\tilde{\alpha}^{\,\ell + \Delta s}}\,, \nonumber
\end{align}
up to corrections suppressed by powers of $1/s_1$. Then, the discontinuity of the amplitude follows by summing over all such transition processes. Most importantly, macroscopic spherical symmetry requires that the probability for a partial wave absorption of $\ell$ be proportional to the $\ell$-th spinning Legendre polynomial. Making this dependence manifest involves expressing our result in a basis which contains the spinning Legendre polynomial, alongside unwanted angular structures whose coefficients vanish. A particularly convenient basis that allows for this involves using the Kravchuk polynomials~\cite{Koekoek:2010}; see Appendix~\ref{sec:spin-universlity} for more details. The outcome of this analysis is that the modulus squared of the coupling of each transition is uniquely fixed to be
\begin{equation}
\label{universal_s1}
     |g_{s_1,s_1+\Delta s, h}^{(\ell)}|^2 = |g_h^{(\ell)}|^2\binom{2\ell}{\ell +|\Delta s|}\,,
\end{equation} 
where all $2\ell +1$ couplings for a fixed $s_1$ state are controlled by the single coupling. Performing the same analysis for the emission process yields the universality for a fixed $s_2$ state:
\begin{align}
\label{universal_s2}
|g_{s_2-\Delta s,s_2, h}^{(\ell)}|^2 = |g_h^{(\ell)}|^2\binom{2\ell}{\ell +|\Delta s|}\,.
\end{align}
Importantly, the couplings in \eqref{universal_s1} and \eqref{universal_s2} must be the same because they share the same process. The same discussion also follows for all values of $s_1$ and $s_2$ within the slow-rotating limit, such that $|g_h^{(\ell)}|^2$ is independent of $s_1$ or $s_2$. This provides a direct emergence of spin universality~\cite{Holstein:2008sw,Holstein:2008sx,Siemonsen:2019dsu}: enforcing consistency with macroscopic symmetry requires that the couplings for all spinning states are proportional to a single universal coupling.

It is further expected that, for a slow-rotating object, the spectrum is insensitive to spin, so that $\rho(M_2,s_2) \simeq \rho(M_2)$. Therefore, we obtain
\begin{align}
    ~\frac{{\rm Disc} \mathcal{A}_{1'\leftarrow 1}^{-h,h}}{2\pi i }&\simeq e^{-\| \alpha\|^2} \sum_{s, \ell} 
 \frac{\rho(M_2)}{(2s)!} \|\alpha \|^{4s} |g^{(\ell)}_{h}|^2 \frac{2\ell +1}{4\pi} \mathcal{P}_{h',h}^{(\ell)}
 \nn
 &= \sum_{\ell} \rho(M_2) |g^{(\ell)}_{h}|^2 \frac{2\ell +1}{4\pi} \mathcal{P}_{h',h}^{(\ell)}\,,
\end{align}
where the amplitudes exponentiate by virtue of spin universality, thereby canceling the otherwise vanishing normalization factor $e^{-\| \alpha \|^2}$ in the classical limit $\|\alpha \|^2 \gg 1$. It then correctly describes the $\mathcal{O}(S^0)$ dynamics. The discontinuity due to the emission is computed in an identical fashion. The probabilities for absorbing and emitting partial-wave $(\ell,m)$ are thus
\begin{align}
    {}_h P^{\rm abs}_{\ell,m}(\omega)&=\frac{|\omega|}{8\pi  M_2 } \rho(M_2) | g^{(\ell)}_h|^2\qquad (\omega>0)
    \,, 
    \label{abs_probability} \\
    {}_{-h} P^{\rm em}_{\ell,m}(\omega)&=\frac{|\omega|}{8\pi M_1 } \rho(M_1) | g^{(\ell)}_h|^2 \qquad (\omega<0)
    \,. \label{em_probability}
\end{align}
The probabilities are controlled by a single coupling, and the only difference between them is the argument of the spectral density. This is Fermi's golden rule, but it has been derived using macroscopic information. Local detailed balance then directly follows from \eqref{abs_probability}, \eqref{em_probability}, and $\rho=Ce^{\mathcal{S}}$ under the assumption that the change of mass is negligible in the polynomial terms:
\begin{align}
    \frac{{}_h P^{\rm abs}_{\ell,m}(\omega)}{{}_{-h} P^{\rm em}_{\ell,m}(-\omega)} = e^{\Delta \mathcal{S}}\,, ~~ \Delta \mathcal{S}=\mathcal{S}(M_2)-\mathcal{S}(M_1)~~~(\omega>0)
    \,.
    \label{fluctuation_th}
\end{align}
We stress that spin universality plays an essential role in this derivation. In particular, the appearance of the same coupling in \eqref{abs_probability} and \eqref{em_probability} does not follow from CPT invariance alone; the probabilities are inclusive, such that ${}_{-h} P^{\rm em}_{\ell,m}$ is not the reverse of ${}_h P^{\rm abs}_{\ell,m}$. Instead, it arises from combining CPT with spin universality~\footnote{For example, the absorption process of a spin-$s$ state means the transition $(M_1,s)$ to $(M_2,s_2)$ with $s_2=s, s\pm 1,\cdots, s \pm \ell$, whereas the emission process is $(M_2,s)$ to $(M_1,s_2)$ with $s_1=s, s\pm 1,\cdots, s \pm \ell$, which is not the reverse process of the absorption. However, the spin universality leads to their transition probabilities being identical up to the spectral density.}.

\textbf{\textit{BHs and extremizing unitarity---}}We now wish to apply our formalism to BHs. Our implementation of the classical absorption properties of BHs involves using the classical transmissivity or the greybody factor $|\mathcal{T}_{\ell}|^2$ as our asymptotic input. The next question is then how we can understand Hawking radiation from a unitary $S$-matrix.

The thermal spectrum is obtained by reconciling quantum probabilities with classical scattering data. In quantum theory, classical waves can be described as coherent states. As detailed in Appendix~\ref{sec:greybody}, provided that the intensity of the classical wave is sufficiently small and that the emission is weak, the classical transmissivity $|\mathcal{T}_{\ell}|^2$ can be expressed in terms of quantum probabilities as
\begin{align}
    |\mathcal{T}_{\ell}(\omega)|^2={}_h P^{\rm abs}_{\ell,m}(\omega) - {}_{-h} P^{\rm em}_{\ell,m}(-\omega) 
    \,. \label{balance_eq}
\end{align}
This equation holds even for strong absorption, $|\mathcal{T}_{\ell}|^2=\mathcal{O}(1)$. Utilizing the conservation law $M_1^2= M_2^2-2M_2\omega$, the spectrum density at $M_1$ can then be approximated by 
$\rho(M_1)\simeq C(M_2)e^{\mathcal{S}(M_2)-\omega/T} $
where $1/T:=\partial \mathcal{S}(M_2)/\partial M_2$.  Importantly, \eqref{abs_probability}, \eqref{em_probability}, and \eqref{balance_eq} determine the coupling to be~\footnote{The frequency $\omega$ here is defined in the rest frame of particle 2, and it can be covariantized as $\omega=p_2\cdot k/M_2$. However, the conservation law imposes $\omega=(M_1^2-M_2^2)/2M_2$, so it is no longer a kinematical variable but a constant in the on-shell 3-point kinematics.}
\begin{align}
    |g^{(\ell)}_h|^2\simeq \frac{8\pi M_2}{\omega C(M_2)e^{\mathcal{S}(M_2)}}\frac{|\mathcal{T}_{\ell}(\omega)|^2e^{\omega/T}}{e^{\omega/T}-1}
    \,.
\end{align}
The emission and absorption probabilities are therefore
\begin{align}
{}_h P^{\rm abs}_{\ell,m}(\omega) &= \frac{|\mathcal{T}_{\ell}|^2e^{\omega/T}}{e^{\omega/T}-1}\,, \ 
{}_
{-h} P^{\rm em}_{\ell,m}(-\omega) = \frac{|\mathcal{T}_{\ell}|^2}{e^{\omega/T}-1}\,,
\label{Hawking_probability}
\end{align}
the latter of which precisely agrees with the thermal spectrum with the greybody factor $|\mathcal{T}_{\ell}|^2$~\footnote{Note that, while \eqref{Hawking_probability} is the quantum probability of one particle absorption/emission, the number of emitted particles is obtained by computing the expectation value of the number operator after time evolution governed by the $S$-matrix. Then, the emission probability \eqref{Hawking_probability} indeed provides the spectrum of emitted particles. See Appendix~\ref{sec:density_matrix} and Ref.~\cite{Aoki:2025ihc} for more details.}.

Our final task is to determine the temperature. 
Classically, a horizon is a boundary of spacetime that maximizes absorption. From the amplitude perspective, this is better phrased as the horizon saturates the maximum absorption {\it allowed} by unitarity. According to unitarity, \eqref{Hawking_probability} leads to the universal upper bound on the greybody factor
\begin{align}
|\mathcal{T}_{\ell}|^2 < 1 - e^{-\omega/T} \,.
\label{T_bound}
\end{align}
It should be noted that this unitarity bound is obtained by neglecting scattering processes other than absorption, realized through a high-energy limit $\omega \gg (\ell+1/2)/r_S$ with $r_S=2GM$~\footnote{Since gravitational interactions inevitably induce additional scattering channels, unitarity should imply a bound stronger than \eqref{T_bound} in general frequencies. In fact, the low-frequency expansion of the greybody factor of the Schwarzschild BH is $|\mathcal{T}_{\ell}|^2=\mathcal{O}(\omega^{2(\ell+1)})$, which is much smaller than \eqref{T_bound}~\cite{Starobinskii:1973hgd,Starobinskil:1974nkd,Page:1976df}. This is understood by the fact that most of the waves cannot be absorbed due to scattering by the potential barrier exterior to the horizon.}. Specifically, for the Schwarzschild BH, the high-energy limit of the greybody factor is
\begin{align}
    |\mathcal{T}_{\ell}|^2\approx 1-e^{-8\pi G M \omega}
    \,,
    \label{greybody_high}
\end{align}
see, e.g., Ref.~\cite{Frolov:1998wf,Raffaelli:2013ih}. 
Therefore, assuming that BHs \eqref{greybody_high} maximize absorption \eqref{T_bound}, we obtain $T=1/8\pi GM$, rediscovering the Hawking temperature.

Note that the high-energy behavior \eqref{greybody_high} is determined by the near-horizon geometry~\cite{Frolov:1998wf,Raffaelli:2013ih}. In other words, \eqref{greybody_high} allows one to extract near-horizon physics directly from the asymptotic data $|\mathcal{T}_{\ell}|^2$. According to Hawking's prediction, the thermal properties of the radiation are likewise governed by this geometry. Crucially, the point of our discussion is not to restate the well-known correspondence between the horizon and Hawking temperature, but rather to highlight the role of unitarity in establishing this correspondence. The unitarity bound is not saturated for BHs in the Boulware vacuum, which may be viewed as the $T\to 0$ limit where the bound reduces to $|\mathcal{T}_{\ell}|^2<1$. As a result, their absence of thermal radiation is not tied to the high-energy greybody behavior \eqref{greybody_high}. More fundamentally, $|\mathcal{T}_{\ell}|^2$ encodes a universal classical property, whereas thermal behavior is determined by the boundary conditions of quantum fields. This observation might suggest a new perspective on BH physics from the standpoint of unitarity: the universality of Hawking radiation---its apparent insensitivity to the BH's internal state and dependence solely on the classical horizon geometry---might arise because these distinct phenomena are connected through the saturation of a unitarity bound.

\textbf{\textit{Discussions---}}In this {\it Letter}, we have studied the absorption and emission of macroscopic objects from a modern on-shell perspective. In a matter of fact, our main results are twofold. First, we have developed a framework for studying non-equilibrium transitions between asymptotic equilibrium states, where spin universality plays a crucial role in describing thermal properties. Second, we have explicitly applied this framework to BHs, finding that the Hawking temperature is associated with maximal absorption under unitary time evolution. This provides a reinterpretation of the correspondence between horizon physics and thermal properties from a unitary $S$-matrix, offering an alternative viewpoint on BH physics.  

Note that we do not claim that Hawking radiation remains universal throughout the entire evaporation process. Rather, our approach identifies the regime in which the semiclassical prediction is expected to break down. Our description is justified only on timescales well below the BH lifetime, in particular before the Page time~\cite{Page:1993df,Page:1993wv}. As a result, we naturally expect deviations from the exact Planck spectrum around the Page time, when BHs can no longer be regarded as asymptotic states and must instead be treated as resonances. Importantly, this opens a promising avenue for investigating the information-loss paradox using modern amplitude program, such as the $S$-matrix bootstrap, where properties of resonances can be extracted from the unitarity and the analyticity of the $S$-matrix~\cite{Hannesdottir:2022bmo,Aoki:2022qbf,Aoki:2023tmu,Correia:2025ozf,He:2023lyy,He:2024nwd,Guerrieri:2024jkn, He:2025gws,Correia:2025uvc,EliasMiro:2026kww}.

Finally, our results tie together several fields, including scattering amplitudes, gravitational waves, BHs, and non-equilibrium phenomena. In particular, it would be interesting to further clarify the connection between our amplitude-based approach and those in non-equilibrium physics, such as the eigenstate thermalization hypothesis~\cite{Deutsch:1991msp,Srednicki:1994mfb,Deutsch:2018ulr,Srednicki:1999bhx,Rigol:2012yjf} and fluctuation theorems, and apply them to BH thermodynamics (see, e.g., Refs.~\cite{Iso:2010tz,Iso:2011gb}). The three-point amplitudes developed here can be glued together using modern on-shell methods to construct higher-point amplitudes, providing a framework for studying the two- and many-body dynamics of thermal objects. Radiation from black holes, neutron stars, and their environments can be systematically incorporated into the contemporary amplitudes program, thereby opening the possibility of testing these ideas through gravitational-wave observations. More broadly, techniques developed in scattering amplitudes and gravitational-wave physics might provide novel tools for understanding BHs and non-equilibrium systems, while insights from these systems may in turn enrich the amplitudes program.

\begin{acknowledgements}

We are grateful to Matteo Sergola and Toshifumi Noumi for early discussions on this project. The work of D.A. was supported in part by The U.K. Royal Society, Grant No.~URF\textbackslash R1\textbackslash 20109, and in part by The Leverhulme Trust, Grant No.~RPG-2024-142. The work of K.A. was supported by JSPS KAKENHI Grant No.~JP24K17046. A.C. was supported by JSPS KAKENHI Grant No. JP24KF0153. The work of H.~J. is supported by the JSPS KAKENHI Grant No.~JP24KJ0902, and World Premier International Research Center Initiative (WPI Initiative), MEXT, Japan. The work of K.Y. is supported by JST SPRING, Grant Number JPMJSP2108 and by a research encouragement grant from the Iwanami Fūjukai. Finally, we acknowledge the 34th edition of the JGRG Conference, held at the Yukawa Institute for Theoretical Physics, and the Komaba Campus of the University of Tokyo, hosted by Toshifumi Noumi, where part of this collaboration was initiated and many fruitful discussions took place. 

\end{acknowledgements}

\bibliography{ref.bib}% Produces the bibliography via BibTeX.

\clearpage
\appendix

\begin{widetext}
\section{Convention}
\label{sec:conventions}
In this appendix we collect our conventions used throughout the paper and the appendices.  The metric signature is $\eta_{\mu\nu}={\rm diag}[+1,-1,-1,-1]$. We absorb factors of $2\pi$ in the integration measure and the delta functions as
\begin{equation}
    \hat{\mathrm{d}}^n p := \frac{\mathrm{d}^n p}{(2\pi)^n}\,, \quad \hat{\delta}^{n}(p) := (2\pi)^n \delta^n(p)\,.
\end{equation}
The Lorentz-invariant phase-space measure is denoted by
\begin{align}
d \Phi(p_1,p_2,\cdots) := \prod_a \hd^4 p_a \hdelta(p_a^2-M_a^2) \theta(p^0_a) = \prod_a \frac{\hd^3 \bm{p}_a}{2E_a} 
\,,
\end{align}
and an on-shell one-particle state is normalized as
\begin{align}
    \bra{p'_1}\ket{p_1}=\hdelta_{\Phi}(p'_1-p_1):=2E_1\hdelta^{(3)}(\bm{p}'_1-\bm{p}_1)
    \,,
\end{align}
with the conventional in-out notation adopted. The free states are denoted by using their labels, and semicolons are used to separate the labels of different particles, e.g., $\ket{p_1,s_1;k}=\ket{p_1,s_1}\otimes \ket{k}$.

In the spinor-helicity formalism, it is customary to use the all-incoming (or all-outgoing) notation. We adopt the all-incoming notation, so $p^0>0$ denotes an incoming particle and $p^0<0$ is an outgoing particle, respectively. This requires flipping the signs of the momentum and the helicity when dealing with outgoing states. One should not confuse the bra-ket notation of the states with the bra-ket of the spinors explained below. The difference between the two should be clear from the context. The Pauli matrices are
\begin{align}
(\sigma^{\mu})_{\alpha \dot{\alpha}}= (\bm{1}, \sigma^i)_{\alpha \dot{\alpha}}\,, \quad
(\bar{\sigma}^{\mu})^{\dot{\alpha}\alpha}=(\bm{1}, -\sigma^i)^{\dot{\alpha}\alpha}
\end{align}
with
\begin{align}
\sigma^1 = 
\begin{pmatrix}
0 & 1 \\
1 & 0 
\end{pmatrix}
\,, \quad
\sigma^2 = 
\begin{pmatrix}
0 & -i \\
i & 0 
\end{pmatrix}
\,, \quad
\sigma^3 =
\begin{pmatrix}
1 & 0 \\
0 & -1 \\
\end{pmatrix}
\,.
\end{align}
The spinor indices are raised and lowered by the Levi-Civita tensor, $\epsilon$, acting from ``left'':
\begin{align}
\epsilon^{\alpha\beta}=\epsilon^{\dot{\alpha}\dot{\beta}}
=
\begin{pmatrix}
0 & 1 \\
-1 & 0
\end{pmatrix}
=-\epsilon_{\alpha\beta}=-\epsilon_{\dot{\alpha}\dot{\beta}}
\,, \quad
\epsilon_{\alpha\beta}\epsilon^{\beta\gamma}=\delta_{\alpha}^{\gamma}
\,.
\end{align}
We adopt the same rules for raising and lowering indices of the $SU(2)$ little group. In the spinor-helicity language, the four-momenta become
\begin{align}
p_{\alpha \dot{\alpha}} &=\sigma^{\mu}_{\alpha \dot{\alpha}} p_{\mu}
\,, \quad p^{\dot{\alpha}\alpha} =p_{\mu}(\bar{\sigma}^{\mu})^{\dot{\alpha}\alpha}\,,
\end{align}
with the explicit massless/massive spinors being
\begin{align}
\text{massless:} \ p_{\alpha \dot{\alpha}}&=\lambda_{\alpha} \tilde{\lambda}_{\dot{\alpha}}
\,, \\
\text{massive:} \ p_{\alpha \dot{\alpha}}&=\lambda^I_{\alpha} \tilde{\lambda}_{I\dot{\alpha}}\,.
\end{align}
We relate spinors with $p$ and $-p$ via
\begin{align}
\lambda(-p)=\lambda(p)\,, \quad \tilde{\lambda}(-p)=-\tilde{\lambda}(p)
\,.
\label{lambda_minus}
\end{align}
For real kinematics, we have
\begin{align}
\text{massless:} \ (\lambda_{\alpha})^* &={\rm sign}(p^0) \tilde{\lambda}_{\dot{\alpha}}\,, \\
\text{massive:} \  (\lambda^I_{\alpha})^*&={\rm sign}(p^0) \tilde{\lambda}_{I\dot{\alpha}} \,, \ 
(\lambda_{I\alpha})^* =-{\rm sign}(p^0) \tilde{\lambda}^I_{\dot{\alpha}}
\,.
\end{align}
In the bra-ket notation, the spinors are represented by
\begin{alignat}{4}
\ket*{p} &\leftrightarrow \lambda_{\alpha}\,, &\quad \bra*{p} &\leftrightarrow \lambda^{\alpha}  \,, &\quad
|p] &\leftrightarrow \tilde{\lambda}^{\dot{\alpha}}\,, &\quad [q| &\leftrightarrow \tilde{\lambda}_{\dot{\alpha}} 
\,,
\\
\ket*{p^I} &\leftrightarrow \lambda^I_{\alpha}\,, &\quad \bra*{p^I} &\leftrightarrow \lambda^{I\alpha}  \,, &\quad
|p^I] &\leftrightarrow \tilde{\lambda}^{I\dot{\alpha}}\,, &\quad [q^I| &\leftrightarrow \tilde{\lambda^I}_{\dot{\alpha}} 
\,.
\end{alignat}
For concrete calculations, we can use the following realizations of massive and massless ingoing spinors:
\begin{align}
\label{spinors}
& \ket*{p_a^I}
 = \sqrt{M_a}\,e^{-\frac{\lambda}{2}(\bm n\cdot\bm\sigma)}\,,
\quad |p_a^I]
= \sqrt{M_a}\,e^{+\frac{\lambda}{2}(\bm n\cdot\bm\sigma)}\,, \\[0.1cm]
& |k\rangle
 = \sqrt{2\omega}
\left(
\begin{smallmatrix}
-e^{-i\varphi/2}\sin\frac{\theta}{2}\\
 e^{ i\varphi/2}\cos\frac{\theta}{2}
\end{smallmatrix}
\right)\,, \ 
|k] = \sqrt{2\omega}
\left(
\begin{smallmatrix}
 e^{-i\varphi/2}\cos\frac{\theta}{2}\\
 e^{ i\varphi/2}\sin\frac{\theta}{2}
\end{smallmatrix}
\right)\,, 
\end{align}
for momenta $p^\mu_a=(M_a \cosh \lambda, M_a \sinh \lambda \bm{n})$ where $\bm{\sigma}$ being the usual three-vector of Pauli matrices, and massless momentum $k^\mu =\omega ( 1, \cos \varphi \sin \theta, \sin \varphi \sin \theta, \cos \theta )$. Here, one can use the formula involving an exponentiation of the Pauli matrices
\begin{align}
e^{i\theta (\boldsymbol{n}\cdot \boldsymbol{\sigma})}= \bm{1} \cos \theta + i \boldsymbol{n}\cdot \boldsymbol{\sigma} \sin \theta 
\,,
\end{align}
with $\theta = \pm \frac{i}{2}\lambda$. Then, choosing the rest frame of either particle 1 or 2, we find
\begin{align}
    \xi_-^I&=\xi_-^J= \left(e^{i\varphi/2}\cos \frac{\theta}{2}, e^{-i\varphi/2}\sin\frac{\theta}{2} \right), \quad
    \xi_+^I=\xi_+^J=\left(e^{i\varphi/2}\sin \frac{\theta}{2}, -e^{-i\varphi/2}\cos\frac{\theta}{2} \right)\,,
\end{align}
and 
\begin{align}
\mathcal{E}^{IJ}&=\begin{pmatrix}
0 & 1 \\
-1 & 0
\end{pmatrix}
\,,\\
{}_h\Ycal \overbrace{{}^{11\cdots 1}}^{\ell-m}\overbrace{{}^{22\cdots 2}}^{\ell + m} &= (-1)^{\ell+h}N\, [{}_{-h}Y_{\ell,m}(\theta,\varphi)]^*
\,, \\
    {}_h \Ycal \underbrace{{}_{11\cdots 1}}_{\ell-m}\underbrace{{}_{22\cdots 2}}_{\ell + m} &= N\, {}_hY_{\ell,m}(\theta,\varphi)
    \,,
\end{align}
with $N=\sqrt{\frac{(\ell+m)!(\ell-m)!}{(\ell!)^2(2\ell)!}}$. 

\section{Thermal radiation from the $S$-matrix}
\label{sec:density_matrix}
Thermal systems are usually described by mixed states. This may seem at odds with our amplitude-based framework, in which thermal radiation arises from transitions between pure states. In this appendix, we clarify the consistency between the two descriptions by discussing how observables restricted to a subsystem are computed from the $S$-matrix, and then demonstrate in a concrete example how thermality emerges from an amplitude-based computation. 

Let us first consider the computation of general observables. We introduce an operator, $\hat{\mathcal{O}}(\alpha)$, acting on a subsystem of the total Hilbert space, $\mathcal{H}=\mathcal{H}_\alpha\otimes\mathcal{H}_\beta$, where $\ket{\alpha}$ denotes the visible sector and $\ket{\beta}$ denotes the invisible sector:
\begin{equation}
    \bra{\alpha';\beta'}\hat{\mathcal{O}}(\alpha)\ket{\alpha;\beta}=\bra{\alpha'}\hat{\mathcal{O}}(\alpha)\ket{\alpha}\delta(\beta-\beta')\,.
\end{equation}
For an initial pure state $\ket{\Psi}$, the observable at future infinity is computed by
\begin{align}
    \mathcal{O}(\alpha)&=\bra{\Psi}\hat{S}^{\dagger}\hat{\mathcal{O}}(\alpha)\hat{S}\ket{\Psi} 
    \nn
     &=\int_{\alpha,\alpha',\beta,\beta'}\bra{\Psi}\hat{S}^{\dagger}\ket{\alpha';\beta'}\bra{\alpha';\beta'}\hat{\mathcal{O}}(\alpha)\ket{\alpha;\beta}\bra{\alpha;\beta}\hat{S}\ket{\Psi} 
    \nn
    &=\int_{\alpha',\alpha} \bra{\alpha'}\hat{\mathcal{O}}(\alpha)\ket{\alpha} P_{\alpha',\alpha}
    \label{Oalpha}
\end{align}
where $\int_{\alpha}$ denotes the on-shell integral and
\begin{align}
     P_{\alpha',\alpha} = \int_{\beta}\bra{\Psi}\hat{S}^{\dagger}\ket{\alpha';\beta} \bra{\alpha;\beta}\hat{S}\ket{\Psi} \,,
\end{align}
is the product of two $S$-matrix elements summed over all invisible $\beta$ states. Its diagonal part $\alpha'=\alpha$ gives the familiar inclusive probabilities. The observable \eqref{Oalpha} can be rewritten compactly as
\begin{align}
        \mathcal{O}(\alpha)={\rm Tr}_{\alpha}(\hat{\mathcal{O}}(\alpha)\hat{\rho}_{\alpha})
    \,,
\end{align}
with
\begin{align}
    \hat{\rho}_{\alpha} &= \int_{\alpha,\alpha'}P_{\alpha,\alpha'}\ket{\alpha'}\bra{\alpha} 
    \label{rho_alpha}\,, 
\end{align}
and ${\rm Tr}_{\alpha}(\hat{A})=\int_{\alpha}\bra{\alpha}\hat{A}\ket{\alpha}$. Therefore, the observable can be computed using the reduced density matrix for the visible sector $\hat{\rho}_{\alpha}$. It is computed by tracing out the invisible sector:
\begin{align}
    \hat{\rho}_{\alpha}={\rm Tr}_{\beta}(\hat{\rho}_{\alpha\beta})\,, \quad
    \hat{\rho}_{\alpha\beta}=\hat{S}\ket{\Psi}\bra{\Psi}\hat{S}^{\dagger}
    \,.
\end{align}
Although the full final state $\hat {S}\ket{\Psi} $ remains pure by unitarity, the reduced states described by $\hat{\rho}_{\alpha}$ are mixed whenever the $S$-matrix entangles these two sectors. Thus, there is no conflict in describing any observables confined to the visible sector between an $S$-matrix description in the full Hilbert space and a density-matrix description in the reduced Hilbert space. 

Having established this general point, we now show how thermal features emerge in amplitude-based calculations. As a concrete example, consider the number operator of the massless spin-0 particle, $\hat{n}_k=\hat{a}_k^\dagger\hat{a}_k$. For an initial
state $\ket{\Psi}$, the expectation value of the number of massless particles radiated to infinity is given by
\begin{align}
    n_k=\bra{\Psi}\hat{S}^{\dagger}\hat{n}_k\hat{S}\ket{\Psi}
    \,,
\end{align}
which can be rewritten in terms of amplitudes by inserting the completeness relations. An important point is that $\hat n_k$ acts only on the massless sector and is therefore blind to the microstate of the massive particle. Thermality emerges precisely from summing over this unresolved internal degeneracy. We thus keep the microstate label $v$ explicit in the following, and write the one-particle completeness relation for the massive sector as
\begin{align}
    \hat{1}=\sum_{s,v}\int\dd \mu^2\int \dd\Phi(p)\ket{p,s,v}\bra{p,s,v}\,.
    \label{completness_micro}
\end{align}
For simplicity, we take the initial state to be a single-particle state with momentum $p_2$ and spin-$s_2$ (strictly speaking, a wave packet state
should be used to describe a localized object, but it does not affect the main disucssion below). We further restrict our analysis to the one-massless particle emission process, which well approximates Hawking radiation as shown in Ref.~\cite{Aoki:2025ihc}. If a system thermalizes after time evolution, the observable $n_k$ should be independent of the concrete realization of the initial microstate. To keep track of the initial state dependence, we write the initial state as 
\begin{align}
    \ket{\Psi}=\sum_{v_2} c_{v_2} \ket{p_2,s_2,v_2}\,, \qquad \sum_{v_2} |c_{v_2}|^2=1
    \,.
\end{align}
Inserting \eqref{completness_micro}, the number of radiated quanta becomes
\begin{align}
    n_k =\sum_{s_1,v_1,v_2,v'_2} \int \dd \mu^2_1 \int \dd \Phi(p_1) c_{v_2^\prime}^* c_{v_2}\bra{p_2,s_2,v'_2}\hat{S}^{\dagger}\ket{p_1,s_1,v_1;k}\bra{p_1,s_1,v_1;k}\hat{S}\ket{p_2,s_2,v_2}
    \,.
    \label{nk_micro}
\end{align}
We now assume that the microstate couplings $g_{v_1,v_2}$ (while suppressing other labels) have (i) equal moduli and (ii) uncorrelated phases, namely
\begin{align}
    g_{v_1,v_2}=e^{i\theta_{v_1,v_2}} g
    \,.
\end{align}
The transition amplitudes between microstates then take the form
\begin{align}
\bra{p_1,s_1,v_1;k}\hat{S}\ket{p_2,s_2,v_2}=e^{i\theta_{v_1,v_2}}\bra{p_1,s_1;k}\hat{S}\ket{p_2,s_2}
\end{align}
where $\ket{p_1,s_1}$ and $\ket{p_2,s_2}$ denote representative states
with coupling $g$. The microstate labels thus enter only through phases, while the modulus is universal. Under these assumptions, and in the limit of a large number of degeneracies, we find
\begin{align}
    n_k &= \sum_{s_1, v_1, v_2} \int \dd \mu^2_1 \int \dd \Phi(p_1) | c_{v_2}|^2\bra{p_2,s_2}\hat{S}^{\dagger}\ket{p_1,s_1;k}\bra{p_1,s_1;k}\hat{S}\ket{p_2,s_2}
    \nn
    &=\sum_{s_1 } \int \dd \mu^2_1 \int \dd \Phi(p_1) \rho(\mu_1,s_1)|\bra{p_1,s_1;k}\hat{S}\ket{p_2,s_2}|^2\,,
\end{align}
where the spectral density $\rho$ counts the number of microstates $v_1$. As a result, the number of radiated quanta---an expectation value in the pure final state $\hat{S}\ket{\Psi}$---is thus computed from the squared three-point amplitude weighted by the spectral density, independently of the initial state $c_v$. Thermal radiation emerges once this inclusive probability takes the form~\eqref{Hawking_probability}, as derived in the main text.

Our assumptions of (i) equal moduli and (ii) uncorrelated phases for the microstate couplings are reminiscent of the eigenstate thermalization hypothesis (ETH)~\cite{Deutsch:1991msp,Srednicki:1994mfb,Deutsch:2018ulr,Srednicki:1999bhx,Rigol:2012yjf}. Assumption (ii) makes the off-diagonal contributions in \eqref{nk_micro} add up as a random phase so that they are exponentially suppressed by $e^{-\mathcal{S}/2}$ relative to the diagonal ones. Assumption (i) further renders the final result independent of the concrete realization of the initial microstates. These two roles parallel the two ETH assumptions described in Ref.~\cite{Rigol:2012yjf}, which constrain the matrix elements of a Hermitian operator so as to relate the observables of one eigenstate to an ensemble average. Note, however, that our assumptions concern the $S$-matrix elements rather than those of a Hermitian operator. They are also closely related to the statements about the kinematic structure of partial-wave absorption and emission. As mentioned in the main text, without assumption (ii), there are interference terms between different spin states, which would lead to unwanted angular dependence of the discontinuities. It would be interesting to explore the interplay of kinematics, macroscopic symmetries, and the ETH from the perspective of on-shell amplitudes.

Before closing this section, it is worth mentioning how the above calculations must be modified if the massive states cannot be treated as asymptotic states. First, one can no longer prepare $\ket{p_2,s_2}$ as an initial state; instead, one must consider a gravitational collapse from stable asymptotic states, e.g., a super-Planckian collision of elementary particles~\cite{Giddings:2007qq, Giddings:2009gj}. This is indeed analogous to Hawking's derivation of thermal radiation~\cite{PhysRevD.14.2460}, where the BH is formed dynamically by gravitational collapse rather than postulated as an eternal asymptotic object. Second, the BH sector must not be traced out as the intermediate cut states. Recall that the observable $\bra{\Psi}\hat{S}^{\dagger}\hat{\mathcal{O}}(\alpha)\hat{S}\ket{\Psi} $ is interpreted as a weighted on-shell cut of the $\Psi \to \Psi$ $S$-matrix element. If the BHs are treated as resonances rather than the asymptotic states, their cuts should expose their decay products~\cite{Veltman:1963th, Denner:2014zga}, not the BH states themselves. These products may contribute to the visible sector, offering an on-shell approach to one of the central questions of quantum gravity: where information resides after BH evaporation.

\section{The computation of probabilities}
\label{sec:probability}
In this appendix we provide a detailed computation of the absorption and emission probabilities. Since the probability of emitting a partial wave $\ell$ and helicity $h$ can be obtained from the probability of absorbing a partial wave $\ell$ and helicity $h$ with the replacement $1\leftrightarrow 2$, $\omega \to -\omega $ and $h\to -h$, we only focus on the absorption probabilities. We start with the $\ell = 0, 1, 2$ cases. Then, by building on these explicit cases, we will derive the result for generic values of $\ell$ presented in the main text. 

\subsection{Partial wave probabilities}
Our starting point is the absorption probability for generic values of $\ell$ in the center-of-mass frame:
\begin{align}
    P_{\ell,m}^{\rm abs}=\sum_{s_2} \int \dd \mu^2 \rho(\mu^2)\int \dd\Phi(p_2) |\bra{p_2,s_2}S\ket{\phi,\gamma}|^2\,,
\end{align}
where the massive particle state and the massless particle state are treated as wavepacket as
\begin{align}
    \ket{\phi}=\int \dd \Phi(p)\phi(p)\ket{p,\alpha}\,,
\end{align}
and
\begin{align}
    \ket{\gamma}&=\int \hd\omega \gamma(\omega)\ket{\omega, \ell,m,h}\,, 
    \\
    \ket{\omega, \ell,m,h} 
    &=\frac{\sqrt{2\omega}}{4\pi} \int \dd^2\hat{\bm{k}}\, {}_{-h}Y_{\ell,m}(\hat{\bm{k}})\ket{k,h}\,,
\end{align}
where $\hat{\bm{k}}=\bm{k}/\omega=(\cos \varphi \sin \theta, \sin \varphi \sin \theta, \cos \theta )$. The normalization of the spherical wave state and the plane wave state are given by 
\begin{align}
    \bra{\omega',\ell',m',h' }\ket{\omega,\ell,m,h}&=\delta_{\ell',\ell}\delta_{m',m}\delta_{h',h}\hat{\delta}(\omega'-\omega)\,,\nonumber\\
    \bra{k',h'}\ket{k,h}&=2\omega \hat{\delta}^3(\bm{k}'-\bm{k})\delta_{h',h}
    \,,
\end{align}
and the wavepacket states are normalized to be unity. The wavepacket for the massive state is chosen to be sharply peaked at the momentum of the wavepacket while it is sufficiently broadened to have a well-defined position; as a concrete realization, for example, we can use a wavepacket~\cite{Kosower:2018adc, Cristofoli:2021vyo}:
\begin{align}
    \phi(p_1)=\mathcal{N} M_1^{-2} \exp\left[ - \frac{p_1 \cdot u}{M_1 \xi}\right],\qquad \frac{\omega^2}{M_1^2}\ll \xi \ll 1\,,
    \label{massive_wp}
\end{align}
where $u$ is the four-velocity of the wavepacket and $\mathcal{N}$ is the normalization factor. This wavepacket satisfies
\begin{align}
    \phi^*(p'_1) \phi(p_1)=|\phi(\bar{p}_1)|^2\,, \qquad \bar{p}_1=\frac{1}{2}(p_1+p'_1)
    \,.
\end{align}
Thus, the wavepacket \eqref{massive_wp} localizes the momentum to be $\bar{p}_1 \propto u_1$, and the absorption probability is
\begin{align}
    P_{\ell,m}^{\rm abs}&=\int \hd \omega \hd \omega' \dd^2\hat{\bm{k}} \dd^2\hat{\bm{k}}' \frac{\sqrt{4\omega \omega'}}{8\pi}
    \hdelta(2\bar{p}_1 \cdot (k-k'))
    {}_{-h}Y^*_{\ell,m}(\hat{\bm{k}}') {}_{-h}Y_{\ell,m}(\hat{\bm{k}})\gamma^*(\omega')\gamma(\omega) \frac{{\rm Disc} \mathcal{A}_{1'\leftarrow 1}^{-h,h}}{2\pi i }
    \nn
    &=\int \hd \omega \hd \omega' \dd^2\hat{\bm{k}} \dd^2\hat{\bm{k}}' \frac{\sqrt{4\omega \omega'}}{8\pi}
    \hdelta(2M_2(\omega-\omega'))
    {}_{-h}Y^*_{\ell,m}(\hat{\bm{k}}') {}_{-h}Y_{\ell,m}(\hat{\bm{k}})\gamma^*(\omega')\gamma(\omega) \frac{{\rm Disc} \mathcal{A}_{1'\leftarrow 1}^{-h,h}}{2\pi i }
    \nn
    &=\frac{|\omega|}{8\pi M_2}\int \dd^2 \hat{\bm{k}} \dd^2 \hat{\bm{k}}'  
\, {}_{-h}Y_{\ell,m}^*(\hat{\bm{k}}') {}_{-h}Y_{\ell,m}(\hat{\bm{k}}) 
~\frac{{\rm Disc} \mathcal{A}_{1'\leftarrow 1}^{-h,h}}{2\pi i }\,,
\end{align}
where
\begin{align}
    {\rm Disc} \mathcal{A}_{1'\leftarrow 1}^{-h,h}\hdelta^{(4)}(p_1+k-p_1'-k')=i\sum_{s_2} \int \dd \mu^2 \rho(\mu^2)\int \dd\Phi(p_2) \bra{p_1',\alpha;k',h}\hat{S}^{\dagger}\ket{p_2,s_2}\bra{p_2,s_2}\hat{S}\ket{p_1,\alpha;k,h}
    \,.
\end{align}
To obtain the second line, we have used $2\bar{p}_1\cdot (k-k')=2p_2\cdot (k-k')$ for $p_2=p_1+k=p_1'+k'$ and $k^2=k'{}^2=0$, and the fact that the frequencies are defined in the center-of-mass frame. Then, we have assumed the massless wavepacket $\gamma(\omega)$ is also localized at a certain frequency, which we rename $\omega$ after performing the integral. Finally, assuming uncorrelated phases of microstate couplings, we obtain
\begin{align}
    P^{\text{abs}}_{\ell,m}&\simeq \frac{|\omega|}{8\pi M_2}\int \dd^2 \hat{\bm{k}} \dd^2 \hat{\bm{k}}'  
\, {}_{-h} Y_{\ell,m}^*(\hat{\bm{k}}') {}_{-h}Y_{\ell,m}(\hat{\bm{k}})e^{-\| \alpha\|^2} \sum_{s_1,s_2,\ell'} 
 \frac{\rho(M_2,s_2)}{(2s_1)!} ||\mathcal{A}^{(\ell')}_{s_1,s_2,h}||^2
 \nn
 &=\frac{|\omega|}{8\pi M_2}\int \dd^2 \hat{\bm{k}} \dd^2 \hat{\bm{k}}'  
\, {}_{-h} Y_{\ell,m}^*(\hat{\bm{k}}') {}_{-h}Y_{\ell,m}(\hat{\bm{k}})e^{-\| \alpha\|^2} \sum_{s_1,s_2} 
 \frac{\rho(M_2,s_2)}{(2s_1)!} ||\mathcal{A}^{(\ell)}_{s_1,s_2,h}||^2
\end{align}
where the second line follows from the orthogonality of the spin-weighted spherical harmonics. The modulus squared of the $\ell$ amplitude reads 
\begin{align}
    ||\mathcal{A}_{s_1, s_2,h}^{(\ell)}||^2 & =  (-1)^{\ell - \Delta s}|g_{s_1, s_2,h}^{(\ell)}|^2  \tilde{\alpha}_{\{ K_{2s_1} \}} {}_{-h}\tilde{\Ycal}^{\{K_1\ldots K_{\ell-\Delta s}}{}_{\{J_1 \ldots J_{\Delta s + \ell}} \ldots \tilde{\mathcal{E}}^{K_{2s_1}\}}{}_{J_{2s_2}\}} \\
    & \hspace{2cm} \times {}_{h}\Ycal_{\{I_1\ldots I_{\ell-\Delta s }}{}^{\{J_1 \ldots J_{\Delta s + \ell}}\ldots \mathcal{E}_{I_{2s_1}\}}{}^{J_{2s_2}\}}\alpha^{\{I_{2s_1}\}}\,,     \label{eq:AsquaredL} \nonumber
\end{align}
and we introduce the notation $\Delta s = s_2 - s_1$. For convenience, we further define the following tensorial objects
\begin{align}
    {}_h \Ycal_{I_1\cdots I_{\ell-\Delta s}}{}^{J_1 \cdots J_{\ell+\Delta s}} &\coloneqq \epsilon_{I_1 K_1} \cdots \epsilon_{I_{\ell - \Delta s} K_{\ell -\Delta s}}\, {}_h \Ycal^{K_1\cdots K_{\ell-\Delta s}J_1 \cdots J_{\ell+\Delta s}}~\,, \\
    {}_h \tilde{\Ycal}_{I_1 I_2\cdots I_{2\ell}} &\coloneqq (-1)^{\ell+h} \, {}_h \Ycal_{I_1 I_2\cdots I_{2\ell}} = \left({}_{-h} \Ycal^{I_1 I_2\cdots I_{2\ell}}\right)^*~\,, 
    \\
    {}_h\tilde{\Ycal}^{I_1\cdots I_{\ell-\Delta s}}{}_{J_1 \cdots J_{\ell+\Delta s}}  &\coloneqq \epsilon^{I_1 K_1} \cdots \epsilon^{I_{\ell - \Delta s} K_{\ell -\Delta s}}\,
    {}_h \tilde{\Ycal}_{K_1\cdots K_{\ell-\Delta s}J_1 \cdots J_{\ell+\Delta s}}=(-1)^{\ell-\Delta s}\left({}_{-h} \Ycal_{I_1\cdots I_{\ell-\Delta s}}{}^{J_1 \cdots J_{\ell+\Delta s}}\right)^*\,,
\end{align}
and
\begin{align}
    \tilde{\mathcal{E}}_{IJ}:=-\mathcal{E}_{IJ}\;,
\end{align}
so that $\tilde{\mathcal{E}}^I{}_J=(\mathcal{E}_I{}^J)^*$, where the equalities relating complex conjugates hold for real kinematics. Throughout our discussion, we will make use of the so-called Schoonschip notation~\cite{Strubbe:1974vj, Veltman:1991xb}
\begin{align}
    {}_h\Ycal_{\alpha^n}{}^{\tilde{\alpha}^{m}}  &  \coloneqq \tilde{\alpha}_{K_1}\ldots \tilde{\alpha}_{K_m} \ {}_h\Ycal_{I_1 \ldots I_n}{}^{K_1 \ldots K_{m}}\alpha^{I_1}\ldots \alpha^{I_n}\;, \\
    {}_h\tilde{\Ycal}^{\tilde{\alpha}^{m}}{}_{\alpha^n}  &  \coloneqq \tilde{\alpha}_{K_1}\ldots \tilde{\alpha}_{K_m} \ {}_h\tilde{\Ycal}^{K_1 \ldots K_{m}}{}_{I_1 \ldots I_n}\alpha^{I_1}\ldots \alpha^{I_n}\;, 
\end{align}
such that in the presence of free indices we have
\begin{align}
   {}_h \Ycal_{\alpha^n}{}^{\tilde{\alpha}^{m-r} J_1 \ldots J_r}  &  \coloneqq \tilde{\alpha}_{K_1}\ldots \tilde{\alpha}_{K_{m-r}} \ {}_h\Ycal_{I_1 \ldots I_n}{}^{K_1 \ldots K_{m-r}  J_1 \ldots J_r}\alpha^{I_1}\ldots \alpha^{I_n}\;, \\
   {}_h \tilde{\Ycal}^{\tilde{\alpha}^m}{}_{\alpha^{n-r}  J_1 \ldots J_r}{} &  \coloneqq \tilde{\alpha}_{K_1}\ldots \tilde{\alpha}_{K_{m}} \ {}_h\tilde{\Ycal}^{K_1 \ldots K_m}{}_{I_1 \ldots I_{n-r}  J_1 \ldots J_r}\alpha^{I_1}\ldots \alpha^{I_{n-r}}\;.
\end{align}
Henceforth, we drop the $h$ subscript in order to reduce clutter, but it should be understood that the results are valid for arbitrary values of $h$. 
\subsection{The $\ell=0$ case}
Before turning to the $\ell>0$ case, let us briefly comment on $\ell=0$. For this value, the only allowed transition is $s_1=s \to s_2=s$, and the corresponding amplitude admits a unique structure
\begin{align}
||\mathcal{A}_{s, s}^{(\ell=0)}||^2 & =  \frac{|g_{s, s}^{(\ell=0)}|^2 }{4\pi} \tilde{\alpha}_{\{ K_{2s_1} \}}  \tilde{\mathcal{E}}^{\{K_1}{}_{\{ J_1}\ldots \tilde{\mathcal{E}}^{K_{2s_1}\}}{}_{J_{2s_2}\}} \times \mathcal{E}{}_{\{J_1}{}^{\{I_1}\ldots \mathcal{E}_{I_{2s_1}\}}{}^{J_{2s_2}\}}\alpha^{\{I_{2s_1}\}}
\nn
&=\frac{|g_{s, s}^{(\ell=0)}|^2}{4\pi} \| \alpha \|^{4s}
\,,
\end{align}
where the factor of $1/4\pi$ has been included to match the normalization used for the $\ell\neq0$ case. Hence, the modulus squared exhibits the desired kinematic structure, and spin universality does not arise from kinematics alone. Instead, spin universality is required to cancel the normalization factor $e^{-\| \alpha\|^2}$ of the spin coherent state:
\begin{align}
     |g_{s, s}^{(\ell=0)}|^2=|g^{(\ell=0)}|^2~~\text{for all } s \implies e^{-\|\alpha \|^2}\sum_s \frac{\rho(M_2)}{(2s)!} ||\mathcal{A}_{s, s}^{(\ell=0)}||^2
    = \frac{1}{4\pi} \rho(M_2) |g^{(\ell=0)}|^2
    \,.
\end{align}
On the other hand, spin universality arises from purely kinematic reasons for $\ell \neq 0$ which is the focus of the remainder of our discussion. 

\subsection{The $\ell=1$ case}
For $\ell = 1$ the modulus squared of the amplitude takes the form
\begin{align}
\label{eq:AsquaredL1}
    ||\mathcal{A}_{s_1, s_2}^{(\ell = 1)}||^2 & = (-1)^{1 - \Delta s} |g_{s_1, s_2}^{(\ell = 1)}|^2 \tilde{\alpha}_{\{ K_{2s_1} \}} \tilde{\Ycal}^{\{K_1\ldots K_{1-\Delta s}}{}_{\{J_1 \ldots J_{\Delta s + 1}} \ldots \tilde{\mathcal{E}}^{K_{2s_1}\}}{}_{J_{2s_2}\}}\\
    & \hspace{3cm} \times \Ycal_{\{I_1\ldots I_{1-\Delta s }}{}^{\{J_1 \ldots J_{\Delta s + 1}}\ldots \mathcal{E}_{I_{2s_1}\}}{}^{J_{2s_2}\}}\alpha^{\{I_{2s_1}\}} \nonumber\,.
\end{align}
As discussed in the main text, the possible transitions for this value of angular momentum are $\Delta s = -1, 0, 1$. In other words, the only allowed absorption processes (and thus emission by CPT) are $s_1 = s \leftrightarrow s_2 = s - 1$, $s_1 = s \leftrightarrow s_2 = s$, and $s_1 = s \leftrightarrow s_2 = s + 1$. Since each process leads to a different structure of \eqref{eq:AsquaredL1}, we will consider each one in turn. 

The first process is $s_1 = s \leftrightarrow s_2 = s - 1$ for which \eqref{eq:AsquaredL1} simply evaluates to
\begin{align}
     ||\mathcal{A}_{s_1, s_2}^{(\ell = 1)}||^2\Big\rvert_{\Delta s = -1} & = |g_{s, s - 1}^{(\ell = 1)}|^2 \tilde{\alpha}_{\{ K_{2s} \}}\tilde{\Ycal}^{\{K_1 K_2} \tilde{\mathcal{E}}^{K_3}{}_{\{J_1} \ldots \tilde{\mathcal{E}}^{K_{2s}\}}{}_{J_{2s - 2}\}} \Ycal_{\{I_1 I_2} \mathcal{E}_{I_3}{}^{\{J_1}\ldots  \mathcal{E}_{I_{2s}\}}{}^{J_{2s-2}\}}\alpha^{\{I_{2s}\}} \\
     & =  |g_{s, s - 1}^{(\ell = 1)}|^2 ||\alpha||^{4s - 4} \tilde{\Ycal}^{\tilde{\alpha} \tilde{\alpha}}\Ycal_{\alpha \alpha}\,. \nonumber 
\end{align}
The second process is $s_1 = s \leftrightarrow s_2 = s$ and yields
\begin{align}
     ||\mathcal{A}_{s_1, s_2}^{(\ell = 1)}||^2\Big\rvert_{\Delta s = 0} & = -|g_{s, s}^{(\ell = 1)}|^2 \tilde{\alpha}_{\{ K_{2s} \}}\tilde{\Ycal}^{\{K_1}{}_{\{J_1} \tilde{\mathcal{E}}^{K_2}{}_{J_2} \ldots \tilde{\mathcal{E}}^{K_{2s}\}}{}_{J_{2s}\}} \Ycal_{\{I_1 }{}^{\{J_1} \mathcal{E}_{I_2}{}^{J_2}\ldots  \mathcal{E}_{I_{2s}\}}{}^{J_{2s}\}}\alpha^{\{I_{2s}\}} \\
     & =  -\frac{|g_{s, s}^{(\ell = 1)}|^2 ||\alpha||^{4s - 4}}{2s} \left(||\alpha||^{2} \tilde{\Ycal}^{\tilde{\alpha}}{}_J \Ycal_{\alpha}{}^{J} + (2s-1)\tilde{\Ycal}^{\tilde{\alpha}}{}_{\alpha }\Ycal_{\alpha}{}^{\tilde{\alpha}}\right)\,, \nonumber 
\end{align}
where the numerical prefactors are purely combinatorial. The final process is  $s_1 = s \leftrightarrow s_2 = s + 1$ and requires a little more care. In this case, \eqref{eq:AsquaredL1} reads
\begin{align}
     ||\mathcal{A}_{s_1, s_2}^{(\ell = 1)}||^2\Big\rvert_{\Delta s = 1} & = |g_{s, s + 1}^{(\ell = 1)}|^2 \tilde{\alpha}_{\{ K_{2s} \}}\tilde{\Ycal}_{\{J_1 J_2} \tilde{\mathcal{E}}^{\{K_1}{}_{J_3} \ldots \tilde{\mathcal{E}}^{K_{2s}\}}{}_{J_{2s  + 2}\}} \Ycal^{\{J_1 J_2} \mathcal{E}_{\{I_1}{}^{J_3}\ldots  \mathcal{E}_{I_{2s}\}}{}^{J_{2s + 2}\}}\alpha^{\{I_{2s}\}} \\
     & = |g_{s, s + 1}^{(\ell = 1)}|^2 \tilde{\alpha}_{\{J_1}\ldots \tilde{\alpha}_{J_{2s}} \tilde{\Ycal}_{J_{2s + 1} J_{2s + 2}\}} \alpha^{\{J_1}\ldots \alpha^{J_{2s}}\Ycal^{J_{2s + 1} J_{2s + 2}\}}\,. \nonumber
\end{align}
To actually go ahead and evaluate this, we will need to think a little deeper about combinatorics. The symmetrized products are expanded as
\begin{equation}
    \alpha^{\{J_1}\ldots \alpha^{J_{2s}}\Ycal^{J_{2s + 1} J_{2s + 2}\}} = \frac{2}{(2s+2)(2s+1)}\left( \alpha^{J_1}\ldots \alpha^{J_{2s}}\Ycal^{J_{2s + 1} J_{2s + 2}} + \text{permutations} \right)\,,
\end{equation}
where the prefactor comes from averaging over the possible positions of the two indices of $\Ycal$. If we fix the positions of the indices on $\Ycal$, then there are three distinct cases for how the indices of $\tilde{\Ycal}$ can be placed with respect to them. The first case is when the indices of $\tilde{\Ycal}$ never contract with $\Ycal$, which yields
\begin{align}
    \tilde{\alpha}_{\{J_1}\ldots \tilde{\alpha}_{J_{2s}} \tilde{\Ycal}_{J_{2s + 1} J_{2s + 2}\}} \alpha^{\{J_1}\ldots \alpha^{J_{2s}}\Ycal^{J_{2s + 1} J_{2s + 2}\}} \supset \frac{2s(2s-1)}{2}||\alpha||^{4s - 4} \tilde{\Ycal}_{\alpha \alpha} \Ycal^{\tilde{\alpha}\tilde{\alpha}}\,. 
\end{align}
The second case is when only one index of $\tilde{\Ycal}$ contracts with $\Ycal$
\begin{align}
    \tilde{\alpha}_{\{J_1}\ldots \tilde{\alpha}_{J_{2s}} \tilde{\Ycal}_{J_{2s + 1} J_{2s + 2}\}} \alpha^{\{J_1}\ldots \alpha^{J_{2s}}\Ycal^{J_{2s + 1} J_{2s + 2}\}} \supset 2(2s)||\alpha||^{4s - 2} \tilde{\Ycal}_{\alpha J} \Ycal^{\tilde{\alpha}J}\,. 
\end{align}
Finally, the third case is when both indices of $\tilde{\Ycal}$ contract with $\Ycal$
\begin{align}
    \tilde{\alpha}_{\{J_1}\ldots \tilde{\alpha}_{J_{2s}} \tilde{\Ycal}_{J_{2s + 1} J_{2s + 2}\}} \alpha^{\{J_1}\ldots \alpha^{J_{2s}}\Ycal^{J_{2s + 1} J_{2s + 2}\}} \supset ||\alpha||^{4s} \tilde{\Ycal}_{J_1 J_2} \Ycal^{J_1 J_2}\,.
\end{align}
Putting all of these cases together, we find 
\begin{align}
     ||\mathcal{A}_{s_1, s_2}^{(\ell = 1)}||^2\Big\rvert_{\Delta s = 1} & = \frac{|g_{s, s + 1}^{(\ell = 1)}|^2 ||\alpha||^{4s - 4}}{(2s + 2)(2s + 1)}\Big( 2||\alpha||^4 \tilde{\Ycal}_{J_1 J_2} \Ycal^{J_1 J_2}\ + 4 (2s) ||\alpha||^2 \tilde{\Ycal}_{\alpha J} \Ycal^{\tilde{\alpha}J} + 2s(2s-1)\tilde{\Ycal}_{\alpha \alpha} \Ycal^{\tilde{\alpha}\tilde{\alpha}} \Big)\,. \nonumber
\end{align}
As we are interested in a classical slow-rotating BH, for which we have $s \gg 1$, the results are simplified to
\begin{align}
    & ||\mathcal{A}_{s_1, s_2}^{(\ell = 1)}||^2\Big\rvert_{\Delta s = 0} \simeq  -|g_{s, s}^{(\ell = 1)}|^2 ||\alpha||^{4s - 4}\tilde{\Ycal}^{\tilde{\alpha}}{}_{\alpha }\Ycal_{\alpha}{}^{\tilde{\alpha}}\,, \quad ||\mathcal{A}_{s_1, s_2}^{(\ell = 1)}||^2\Big\rvert_{\Delta s = 1} \simeq |g_{s, s+1}^{(\ell = 1)}|^2 ||\alpha||^{4s - 4}\tilde{\Ycal}_{\alpha \alpha} \Ycal^{\tilde{\alpha}\tilde{\alpha}}\,.
\end{align}
Finally, summing over all transition processes yields the transition-inclusive modulus squared
\begin{equation}
    ||\mathcal{A}_{s}^{(\ell = 1)}||^2 =  ||\alpha||^{4s - 4}\Big(|g_{s, s - 1}^{(\ell = 1)}|^2 \tilde{\Ycal}^{\tilde{\alpha} \tilde{\alpha}}\Ycal_{\alpha \alpha} - |g_{s, s}^{(\ell = 1)}|^2 \tilde{\Ycal}^{\tilde{\alpha}}{}_{\alpha }\Ycal_{\alpha}{}^{\tilde{\alpha}} + |g_{s, s+1}^{(\ell = 1)}|^2 \tilde{\Ycal}_{\alpha \alpha} \Ycal^{\tilde{\alpha}\tilde{\alpha}} \Big)\,.
\end{equation}

\subsection{The $\ell=2$ case} 

In the case of $\ell = 2$, the modulus squared of the amplitude takes the form
\begin{align}
    ||\mathcal{A}_{s_1, s_2}^{(\ell = 2)}||^2 & = 
    (-1)^{2-\Delta s}|g_{s_1, s_2}^{(\ell = 2)}|^2 \tilde{\alpha}_{\{ K_{2s_1} \}} \tilde{\Ycal}^{\{K_1\ldots K_{2-\Delta s}}{}_{\{J_1 \ldots J_{\Delta s + 2}} \ldots \tilde{\mathcal{E}}^{K_{2s_1}\}}{}_{J_{2s_2}\}} \\
    & \hspace{3cm} \times \Ycal_{\{I_1\ldots I_{2-\Delta s }}{}^{\{J_1 \ldots J_{\Delta s + 2}}\ldots \mathcal{E}_{I_{2s_1}\}}{}^{J_{2s_2}\}}\alpha^{\{I_{2s_1}\}} \nonumber \,,
    \label{eq:AsquaredL2}
\end{align}
for which there are five possible transitions $-2 \leq \Delta s \leq 2$. For each given transition, the amplitudes take the form
\begin{align}
    s_1 = s \leftrightarrow s_2 = s-2:&~\Ycal_{\left\{I_1I_2I_3I_4\right.}\mathcal{E}_{I_5}{}^{\left\{J_1\right.}\cdots \mathcal{E}_{\left.I_{2s}\right\}}{^{\left.J_{2s-4}\right\}}}\;,\nonumber\\
    s_1 = s \leftrightarrow s_2 = s-1:&~\Ycal_{\left\{I_1I_2I_3\right.}{}^{\left\{J_1\right.}\mathcal{E}_{I_4}{}^{J_2}\cdots \mathcal{E}_{\left.I_{2s}\right\}}{^{\left.J_{2s-2}\right\}}}\;,\nonumber\\
    s_1 = s \leftrightarrow s_2 = s:&~\Ycal_{\left\{I_1I_2\right.}{}^{\left\{J_1J_2\right.}\mathcal{E}_{I_3}{}^{J_3}\cdots \mathcal{E}_{\left.I_{2s}\right\}}{^{\left.J_{2s}\right\}}}\;,\nonumber\\
    s_1 = s \leftrightarrow s_2 = s+1:&~\Ycal_{\left\{I_1\right.}{}^{\left\{J_1J_2J_3\right.}\mathcal{E}_{I_2}{}^{J_4}\cdots \mathcal{E}_{\left.I_{2s}\right\}}{^{\left.J_{2s+2}\right\}}}\;,\nonumber\\
    s_1 = s \leftrightarrow s_2 = s+2:&~\Ycal^{\left\{J_1J_2J_3J_4\right.}\mathcal{E}_{\left\{I_1\right.}{}^{J_5}\cdots \mathcal{E}_{\left.I_{2s}\right\}}{^{\left.J_{2s+4}\right\}}}\;.\nonumber
\end{align}
As before, we simply need to compute each case separately, keeping in mind that the combinatorial factors follow straightforwardly from our previous discussion. To this end, each transition yields
\begin{align}
    ||\mathcal{A}_{s_1, s_2}^{(\ell = 2)}||^2\Big\rvert_{\Delta s = -2} & = |g_{s, s-2}^{(\ell = 2)}|^2 \tilde{\alpha}_{\{ K_{2s} \}} \tilde{\Ycal}^{\{K_1K_2K_3K_4}\tilde{\mathcal{E}}^{K_{5}}{}_{\{J_{1}} \ldots \tilde{\mathcal{E}}^{K_{2s}\}}{}_{J_{2s - 4}\}}\Ycal_{\{I_1I_2 I_3 I_4}  \mathcal{E}_{I_{5}}{}^{\{J_{1}}\ldots \mathcal{E}_{I_{2s}\}}{}^{J_{2s - 4}\}}\alpha^{\{I_{2s}\}} \\
    & =|g_{s, s-2}^{(\ell = 2)}|^2 ||\alpha||^{4s - 8} \tilde{\Ycal}^{\tilde{\alpha}\tilde{\alpha}\tilde{\alpha}\tilde{\alpha}}\Ycal_{\alpha \alpha \alpha \alpha}\,, \nonumber
\end{align}
\begin{align}
    ||\mathcal{A}_{s_1, s_2}^{(\ell = 2)}||^2\Big\rvert_{\Delta s = -1} & = -|g_{s, s-1}^{(\ell = 2)}|^2 \tilde{\alpha}_{\{ K_{2s} \}} \tilde{\Ycal}^{\{K_1K_2K_3}{}_{\{J_1}\tilde{\mathcal{E}}^{K_{4}}{}_{J_{2}} \ldots \tilde{\mathcal{E}}^{K_{2s}\}}{}_{J_{2s - 2}\}}\Ycal_{\{I_1I_2 I_3}{}^{\{J_1}  \mathcal{E}_{I_{4}}{}^{J_{2}}\ldots \mathcal{E}_{I_{2s}\}}{}^{J_{2s - 2}\}}\alpha^{\{I_{2s}\}} \\
    & = -\frac{|g_{s, s-1}^{(\ell = 2)}|^2 ||\alpha||^{4s - 8}}{2s-2} \Big(||\alpha||^2 \tilde{\Ycal}^{\tilde{\alpha}\tilde{\alpha}\tilde{\alpha}}{}_J \Ycal_{\alpha \alpha \alpha}{}^J +(2s-3)\tilde{\Ycal}^{\tilde{\alpha}\tilde{\alpha}\tilde{\alpha} }{}_{\alpha}\Ycal_{\alpha \alpha \alpha}{}^{ \tilde{\alpha}} \Big)\,, \nonumber
\end{align}
\begin{align}
    ||\mathcal{A}_{s_1, s_2}^{(\ell = 2)}||^2\Big\rvert_{\Delta s = 0} & = |g_{s, s}^{(\ell = 2)}|^2 \tilde{\alpha}_{\{ K_{2s} \}} \tilde{\Ycal}^{\{K_1K_2}{}_{\{J_1 J_2}\tilde{\mathcal{E}}^{K_{3}}{}_{J_{3}} \ldots \tilde{\mathcal{E}}^{K_{2s}\}}{}_{J_{2s}\}}\Ycal_{\{I_1I_2 }{}^{\{J_1 J_2}  \mathcal{E}_{I_{3}}{}^{J_{3}}\ldots \mathcal{E}_{I_{2s}\}}{}^{J_{2s}\}}\alpha^{\{I_{2s}\}} \\
    & = \frac{|g_{s, s}^{(\ell = 2)}|^2 ||\alpha||^{4s - 8}}{2s(2s-1)} \Big(2||\alpha||^4 \tilde{\Ycal}^{\tilde{\alpha}\tilde{\alpha}}{}_{J_1 J_2}\Ycal_{\alpha \alpha}{}^{J_1 J_2} + 4(2s-2)||\alpha||^2 \tilde{\Ycal}^{\tilde{\alpha}\tilde{\alpha}}{}_{\alpha J}\Ycal_{\alpha \alpha}{}^{ \tilde{\alpha}J} 
    \nonumber \\
    &\qquad \qquad\qquad \qquad\quad + (2s-2)(2s-3)\tilde{\Ycal}^{\tilde{\alpha} \tilde{\alpha} }{}_{\alpha \alpha}\Ycal_{\alpha \alpha }{}^{\tilde{\alpha} \tilde{\alpha}} \Big)\,, \nonumber
\end{align}
\begin{align}
    ||\mathcal{A}_{s_1, s_2}^{(\ell = 2)}||^2\Big\rvert_{\Delta s = 1} & = -|g_{s, s + 1}^{(\ell = 2)}|^2 \tilde{\alpha}_{\{ K_{2s} \}} \tilde{\Ycal}^{\{K_1}{}_{\{J_1 J_2 J_3}\tilde{\mathcal{E}}^{K_{2}}{}_{J_{4}} \ldots \tilde{\mathcal{E}}^{K_{2s}\}}{}_{J_{2s + 2}\}}\Ycal_{\{I_1}{}^{\{J_1 J_2 J_3}  \mathcal{E}_{I_{2}}{}^{J_{4}}\ldots \mathcal{E}_{I_{2s}\}}{}^{J_{2s + 2}\}}\alpha^{\{I_{2s}\}} \\
    & =- \frac{|g_{s, s + 1}^{(\ell = 2)}|^2 ||\alpha||^{4s - 8}}{2s(2s+1)(2s+2)} \Big(6||\alpha||^6 \tilde{\Ycal}^{\tilde{\alpha}}{}_{J_1 J_2 J_3}\Ycal_{\alpha }{}^{J_1 J_2 J_3} + 18(2s-1)||\alpha||^4 \tilde{\Ycal}^{\tilde{\alpha} }{}_{\alpha J_1 J_2}\Ycal_{\alpha }{}^{ \tilde{\alpha} J_1 J_2} \nonumber \\ 
    & \hspace{2cm} + 9 (2s-1)(2s-2)||\alpha||^2\tilde{\Ycal}^{\tilde{\alpha} }{}_{\alpha  \alpha J }\Ycal_{\alpha }{}^{\tilde{\alpha}  \tilde{\alpha}J} + (2s-1)(2s-2)(2s-3)\tilde{\Ycal}^{\tilde{\alpha}}{}_{\alpha  \alpha \alpha}\Ycal_{\alpha}{}^{\tilde{\alpha}  \tilde{\alpha} \tilde{\alpha}} \Big)\,, \nonumber
\end{align}
\begin{align}
    ||\mathcal{A}_{s_1, s_2}^{(\ell = 2)}||^2\Big\rvert_{\Delta s = 2} & = |g_{s, s + 2}^{(\ell = 2)}|^2 \tilde{\alpha}_{\{ K_{2s} \}} \tilde{\Ycal}_{\{J_1 J_2 J_3 J_4}\tilde{\mathcal{E}}^{\{K_{1}}{}_{J_{5}} \ldots \tilde{\mathcal{E}}^{K_{2s}\}}{}_{J_{2s + 4}\}}\Ycal^{\{J_1 J_2 J_3 J_4}  \mathcal{E}_{\{I_{1}}{}^{J_{5}}\ldots \mathcal{E}_{I_{2s}\}}{}^{J_{2s + 4}\}}\alpha^{\{I_{2s}\}} \\
    & = \frac{|g_{s, s + 2}^{(\ell = 2)}|^2 ||\alpha||^{4s - 8}}{(2s+1)(2s+2)(2s+3)(2s+4)} \Big(24 ||\alpha||^8 \tilde{\Ycal}_{J_1 J_2 J_3 J_4}\Ycal^{J_1 J_2 J_3 J_4} + 96(2s)||\alpha||^6\tilde{\Ycal}_{\alpha J_1 J_2 J_3}\Ycal^{\tilde{\alpha}  J_1 J_2 J_3} \nonumber \\
    & \hspace{2cm} +72(2s)(2s-1)||\alpha||^4\tilde{\Ycal}_{\alpha \alpha J_1 J_2 }\Ycal^{\tilde{\alpha} \tilde{\alpha} J_1 J_2} + 16 (2s) (2s-1)(2s-2)||\alpha||^2 \tilde{\Ycal}_{\alpha \alpha \alpha J}\Ycal^{\tilde{\alpha} \tilde{\alpha}\tilde{\alpha} J} 
    \nonumber \\
    & \hspace{2cm} + (2s)(2s-1)(2s-2)(2s-3)\tilde{\Ycal}_{\alpha \alpha \alpha \alpha}\Ycal^{\tilde{\alpha} \tilde{\alpha}\tilde{\alpha}\tilde{\alpha}} \Big)\,. \nonumber
\end{align}
Then, for a classical slow-rotating BH, only the leading structures proportional to $||\alpha||^{4s - 8}$ contribute, such that 
\begin{align}
    ||\mathcal{A}_{s}^{(\ell = 2)}||^2 = ||\alpha||^{4s - 8}\Big(&|g_{s, s - 2}^{(\ell = 2)}|^2 \tilde{\Ycal}^{\tilde{\alpha} \tilde{\alpha} \tilde{\alpha} \tilde{\alpha}}\Ycal_{\alpha \alpha  \alpha  \alpha} - |g_{s, s-1}^{(\ell = 2)}|^2 \tilde{\Ycal}^{\tilde{\alpha} \tilde{\alpha} \tilde{\alpha}}{}_{\alpha}\Ycal_{\alpha \alpha  \alpha}{}^{\tilde{\alpha}} + |g_{s, s}^{(\ell = 2)}|^2 \tilde{\Ycal}^{\tilde{\alpha} \tilde{\alpha}}{}_{\alpha \alpha}\Ycal_{\alpha \alpha}{}^{\tilde{\alpha} \tilde{\alpha}} \\
    & - |g_{s, s+1}^{(\ell = 2)}|^2 \tilde{\Ycal}^{\tilde{\alpha}}{}_{ \alpha \alpha \alpha}\Ycal_{\alpha}{}^{ \tilde{\alpha}  \tilde{\alpha} \tilde{\alpha}} + |g_{s, s+2}^{(\ell = 2)}|^2 \tilde{\Ycal}_{\alpha\alpha \alpha \alpha}\Ycal^{\tilde{\alpha}  \tilde{\alpha}  \tilde{\alpha} \tilde{\alpha}}\Big)\,. \nonumber
\end{align}

\subsection{The generic $\ell$ case}

Given our analysis for the $\ell = 1$ and $\ell = 2$ cases, it is straightforward to extend the result to generic values of $\ell$ by identifying the underlying pattern. Towards this end, we take $s_1 = s$ and $s_2 = s + \Delta s$ with the possible transitions controlled by $- \ell \leq \Delta s \leq \ell$. Note that for a given value of $\Delta s $, the amplitude schematically takes the form
\begin{equation}
    \Ycal_{\{I_1 \ldots I_p}{}^{\{J_1 \ldots J_q} \mathcal{E}_{I_{p+1}}{}^{J_{q + 1}}\ldots \mathcal{E}_{I_{2s}\}}{}^{J_{2s + 2\Delta s}\}}\;,
\end{equation}
where $p = \ell - \Delta s $ and $q = \ell + \Delta s$. Note that, in our convention for $\tilde{\Ycal}$, rewriting the complex conjugate of a mixed-index tensor with $p$ lowered indices in terms of $\tilde{\Ycal}$ produces a factor of $(-1)^p=(-1)^{\ell-\Delta s}$. Once we have contracted everything with $\tilde{\alpha}$ and $\alpha$ when computing the modulus squared of the amplitude, the remaining combinatorial factors depend on how the $q$ $J$-indices of $\tilde{\Ycal}$ overlap with the $q$ $J$-indices of $\Ycal$. One can then conclude that for a given $\Delta s $ the modulus squared of the amplitude takes the form
\begin{equation}
     ||\mathcal{A}_{s_1, s_2}^{(\ell )}||^2\Big\rvert_{\Delta s } = (-1)^p|g_{s, s+\Delta s}^{(\ell)}|^2 \sum_{r = 0}^{q} \frac{\binom{q}{r} \binom{n_s - q}{q-r}}{\binom{n_s}{q}} ||\alpha||^{2n_s - 4q + 2r}\tilde{\Ycal}^{\tilde{\alpha}^{\,p}}{}_{\alpha^{\,q-r}J_1\ldots J_r}\Ycal_{\alpha^{\,p}}{}^{\tilde{\alpha}^{\,q-r}J_1\ldots J_r}\,,
\end{equation}
with $n_s = 2s + 2 \Delta s$, and where $r$ is the number of overlapping indices. Because we are ultimately interested in a classical slow-rotating object, we then take the limit $s \gg 1$. In this limit, we have that the combinatorial prefactors behave as $\mathcal{O}(s^{-r})$ so that the $ r = 0$ term always dominates. Indeed, noticing that $n_s \sim 2s$ and $q \sim \ell$, the $r = 0$ prefactor becomes unity. Thus, for each transition we have
\begin{equation}
     ||\mathcal{A}_{s_1, s_2}^{(\ell )}||^2\Big\rvert_{\Delta s} \simeq (-1)^{\ell -\Delta s} |g_{s, s+\Delta s}^{(\ell)}|^2 ||\alpha||^{4s - 4\ell}\tilde{\Ycal}^{\tilde{\alpha}^{\,\ell -\Delta s}}{}_{\alpha^{\,\ell + \Delta s}}\Ycal_{\alpha^{\,\ell - \Delta s}}{}^{\tilde{\alpha}^{\,\ell + \Delta s}}\,,
\end{equation}
up to corrections suppressed by $1/s$. Finally, summing over all transitions we obtain 
\begin{equation}
     ||\mathcal{A}_{s}^{(\ell )}||^2 = ||\alpha||^{4s - 4\ell} \sum_{\Delta s = - \ell}^{\ell} (-1)^{\ell -\Delta s} |g_{s, s+\Delta s}^{(\ell)}|^2 \tilde{\Ycal}^{\tilde{\alpha}^{\,\ell -\Delta s}}{}_{\alpha^{\,\ell + \Delta s}}\Ycal_{\alpha^{\,\ell - \Delta s}}{}^{\tilde{\alpha}^{\,\ell + \Delta s}}\,,
\end{equation}
and thus the absorption probability for generic $\ell$ takes the form
%_s
\begin{align}
     P^{\rm abs}_{\ell,m} \simeq  \frac{\omega}{8\pi M_2}\int &\dd^2 \hat{\bm{k}} \dd^2 \hat{\bm{k}}'  
\, Y_{\ell,m}^*(\hat{\bm{k}}') Y_{\ell,m}(\hat{\bm{k}})
\nn
&\times
     \sum_{s} \frac{\rho(M_2)}{(2s)!} e^{-||\alpha||^2} ||\alpha||^{4s - 4\ell} \sum_{\Delta s = - \ell}^{\ell} (-1)^{\ell -\Delta s} |g_{s, s+\Delta s}^{(\ell)}|^2 \tilde{\Ycal}^{\tilde{\alpha}^{\,\ell -\Delta s}}{}_{\alpha^{\,\ell + \Delta s}}\Ycal_{\alpha^{\,\ell - \Delta s}}{}^{\tilde{\alpha}^{\,\ell + \Delta s}}\,.
\end{align}

\section{The emergence of spin universality}
\label{sec:spin-universlity}
In this appendix we provide a discussion of the emergence of spin universality for generic $\ell$ values. For consistency with macroscopic spherical symmetry, the probability that a partial wave of angular momentum $\ell$ is absorbed must be independent of the orientation of the spin vector. This kinematic consistency implies the probability to be proportional to the $\ell$-th spinning Legendre polynomial $\mathcal{P}^{(\ell)}$, as it is only the object without the directional dependence of $\alpha_I$. As we have discussed in the main text, the consequence of this condition is that the modulus squared of each transition coupling for all spinning states is proportional to a single universal coupling, say, $|g^{(\ell)}|^2$, and then the probability becomes independent of the size of the classical spin $\|\alpha\|^2$ as well, correctly describing the $\mathcal{O}(S^0)$ dynamics. Our goal in this appendix is therefore to find the precise relation for all couplings.

Focusing on the approximation in which $s \gg 1$, we have from Appendix~\ref{sec:probability} that 
\begin{equation}
    ||\mathcal{A}_{s}^{(\ell )}||^2 = ||\alpha||^{4s-4\ell}\sum_{q = 0}^{n} G_q^{(\ell)} S_q^{(\ell)}\,,
    \label{eq:AppendixCAmplitudeSquaredL}
\end{equation}
where $q = \ell + \Delta s$, $n = 2\ell$, $G_q^{(\ell)} = |g_{s,s-\ell+q}^{(\ell)}|^2$, and 
\begin{equation}
     S_q^{(\ell)} = (-1)^{q}\tilde{\Ycal}^{\tilde{\alpha}^{\,n-q}}{}_{\alpha^{\,q}}\Ycal_{\alpha^{\,n -q}}{}^{\tilde{\alpha}^{\,q}}\,,
\end{equation}
because $(-1)^{\ell - \Delta s} = (-1)^{n - q} = (-1)^q$. The desired $\mathcal{P}^{(\ell)}$-type angular dependence is obtained by inserting the spinor completeness relation
\begin{equation}
    \delta^{I}{}_{J} = \frac{1}{||\alpha||^2}(\alpha^I \tilde{\alpha}_J - \tilde{\alpha}^I \alpha_J)\,,
\end{equation}
$2\ell$ times into $\tilde{\Ycal}^{J_1 \ldots J_{2\ell}}\Ycal_{J_1 \ldots J_{2\ell}}$. This gives 
\begin{equation}
     \tilde{\mathcal{P}}^{(\ell)} \coloneqq \frac{2\ell +1}{4\pi} \mathcal{P}^{(\ell)} = \tilde{\Ycal}^{J_1 \ldots J_{2\ell}}\Ycal_{J_1 \ldots J_{2\ell}} = \frac{1}{||\alpha||^{4\ell} }\sum_{q = 0}^{n} \binom{n}{q}S_q^{(\ell)}\,,
    \label{eq:LegendrePolyL}
\end{equation}
where we define a normalized spinning Legendre polynomial for convenience, and the binomial vector
\begin{equation}
    \left(\binom{n}{0},\binom{n}{1}, \ldots \binom{n}{n} \right)\,,
\end{equation}
is the coefficient vector corresponding to $\tilde{\mathcal{P}}^{(\ell)}$. Our next step is to express \eqref{eq:AppendixCAmplitudeSquaredL} in terms of a basis that contains $\tilde{\mathcal{P}}^{(\ell)}$ plus some unwanted structures whose coefficients we will ultimately enforce to be zero. Indeed, doing this explicitly for generic values of $\ell$ is an involved process, so we should carefully choose a basis for this purpose. It turns out that the so-called Kravchuk polynomials (see, e.g.,~Ref.~\cite{Koekoek:2010}) provide an exceptionally simple way of doing this. The Kravchuk polynomials $K_m(q)$ for $m = 0, \ldots, n$ are
 defined by
\begin{equation}
    \sum_{m = 0}^{n} K_m(q)t^m = (1-t)^q(1+t)^{n-q}\,,
    \label{eq:KravchukPoly}
\end{equation}
where $t$ is simply an auxiliary bookkeeping variable, whereby the coefficient of each power of $t^m$ defines $K_m(q)$. The Kravchuk polynomials obey the orthogonality relation
\begin{equation}
    \sum_{q = 0}^{n} \binom{n}{q} K_m(q) K_{m'}(q) = 2^n \binom{n}{m} \delta_{mm'}\,.
\end{equation}
Thus, this makes them a natural basis for decomposing coefficient vectors with binomial weighting. 

Given these polynomials, we define a basis given by 
\begin{equation}
    U_m^{(\ell)} \coloneqq \frac{1}{||\alpha||^{4\ell} } \sum_{q = 0}^{n} \binom{n}{q} K_m(k) S_q^{(\ell)}\,,
\end{equation}
where, since $K_0(q) = 1$, the first element is precisely $\tilde{\mathcal{P}}^{(\ell)}$
\begin{equation}
    U_0^{(\ell)} = \frac{1}{||\alpha||^{4\ell} } \sum_{q = 0}^{n} \binom{n}{q} S_q^{(\ell)} = \tilde{\mathcal{P}}^{(\ell)}\,,
\end{equation}
while the remaining $U_m^{(\ell)}$ with $m \geq 1$ are all unwanted angular structures, dependent on the direction of $\alpha_I$. By decomposing the modulus squared of the amplitude summed over all transition processes into this basis
\begin{equation}
   ||\mathcal{A}_{s}^{(\ell )}||^2 = ||\alpha||^{4s-4\ell}\sum_{q = 0}^{n} G_q^{(\ell)} S_q^{(\ell)} = ||\alpha||^{4s} \sum_{m = 0}^{n} c_m U_m^{(\ell)}\,,
    \label{eq:KravchukBasis}
\end{equation}
followed by using the orthogonality of the Kravchuk polynomials, we find
\begin{equation}
   c_m = \frac{1}{2^n  \binom{n}{m}} \sum_{q = 0}^{n} G_q^{(\ell)} K_m(q)\,.
\end{equation}
Specifically, for pure $ \tilde{\mathcal{P}}^{(\ell)}$ dependence we require that $c_m = 0$ for $m \geq 1$, or, equivalently, 
\begin{equation}
    \sum_{q = 0}^{n} G_q^{(\ell)} K_m(q) = 0\,, \quad m\geq 1\,.
\end{equation}
Imposing this at the level of \eqref{eq:KravchukBasis}
\begin{align}
     ||\mathcal{A}_{s}^{(\ell )}||^2 = ||\alpha||^{4s-4\ell}\sum_{q = 0}^{n} G_q^{(\ell)} S_q^{(\ell)} = ||\alpha||^{4s} c_0 U_0^{(\ell)} =  ||\alpha||^{4s - 4\ell}  \sum_{q = 0}^{n} c_0 \binom{n}{q} S_q^{(\ell)}\;,
\end{align}
allows us to read off that
\begin{equation}
    G_k^{(\ell)} = c_0  \binom{n}{q} = |g^{(\ell)}|^2 \binom{n}{q}\,,
\end{equation}
where the coupling of $\tilde{\mathcal{P}}^{(\ell)}$ is chosen to be $c_0 = |g^{(\ell)}|^2$, such that each modulus squared of the coupling is given by 
\begin{equation}
     |g_{s,s+\Delta s}^{(\ell)}|^2 = |g^{(\ell)}|^2 \binom{2\ell}{\ell +\Delta s}\,.
     \label{App_gabs}
\end{equation}
Thus, the modulus of the amplitude squared summed over all transition processes becomes 
\begin{equation}
    ||\mathcal{A}_{s}^{(\ell )}||^2 = ||\alpha||^{4s-4\ell}\sum_{q = 0}^{n} G_q^{(\ell)} S_q^{(\ell)} = ||\alpha||^{4s-4\ell}\sum_{q = 0}^{n}|g^{(\ell)}|^2 \binom{n}{q}S_q^{(\ell)} = ||\alpha||^{4s}|g^{(\ell)}|^2\tilde{\mathcal{P}}^{(\ell)}\,.
\end{equation}
Therefore, the Kravchuk polynomials diagonalize the space of coefficient vectors with binomial weighting. In this basis, $U_0^{(\ell)}= \tilde{\mathcal{P}}^{(\ell)}$, while the $U_m^{(\ell)}$ with $m\geq 1$ correspond to unwanted angular structures.

The inclusive emission probability---the probability of emitting a partial wave $\ell$ from particle 2, summing over all final states of particle 1---is computed by replacing $1\leftrightarrow 2$ and $\omega \to -\omega$. Hence, the consistency with the macroscopic spherical symmetry of emission concludes
\begin{align}
     |g_{s-\Delta s,s}^{(\ell)}|^2 = |g^{(\ell)}|^2 \binom{2\ell}{\ell -\Delta s}\,.
     \label{App_gem}
\end{align}
Here, an identity of the binomial coefficients
\begin{align}
    \binom{2\ell}{\ell +\Delta s} = \binom{2\ell}{\ell -\Delta s}
\end{align}
implies that the coupling squares for the spin change $\Delta s$ and $-\Delta s$ are identical. Note that the absorption and emission processes are not reverse processes. 
\begin{figure}[t]
\centering
\includegraphics[width=0.7\linewidth]{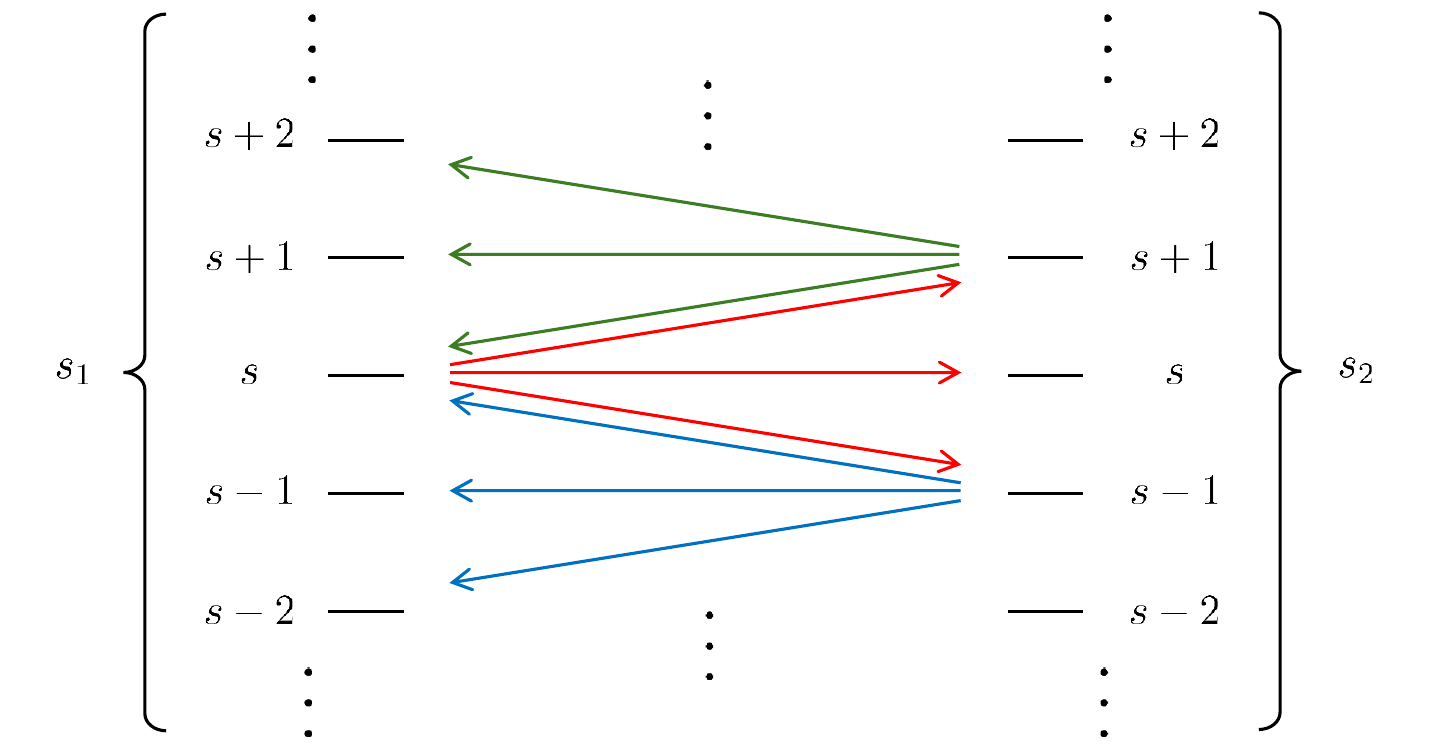}
\caption{Spin transition processes for absorption (left to right) and emission (right to left).}
\label{fig:spin}
\end{figure}
As shown in Fig.~\ref{fig:spin}, an absorption process corresponds to a process from a fixed $s_1$ state into all possible $s_2$ states, and emission is from a fixed $s_2$ to all $s_1$. Macroscopic symmetry has concluded that the $2\ell +1$ couplings for a fixed $s_1$ or $s_2$ are controlled by the same coupling. However, as clearly seen in Fig.~\ref{fig:spin}, the absorption and emission processes can share the same spin transition, for instance, $s_1=s \leftrightarrow s_2=s+1$ and $s_1=s \leftrightarrow s_2=s-1$. Therefore, the microscopic CPT invariance, implemented in crossing symmetry of amplitudes, connects the couplings in \eqref{App_gabs} and \eqref{App_gem}, concluding that the couplings of \emph{all} spin transition processes are controlled by a single universal coupling. In short, spin universality emerges from reconciling the microscopic crossing symmetry and the macroscopic symmetry. This is particularly interesting for photons and gravitons, where they have no exceptional $\ell=0$ process.

Having obtained spin universality, the absorption probability takes the final form
\begin{align}
     P^{\rm abs}_{\ell,m} &\simeq \frac{|\omega|}{8\pi M_2}\int \dd^2 \hat{\bm{k}} \dd^2 \hat{\bm{k}}'  
\, Y_{\ell,m}^*(\hat{\bm{k}}') Y_{\ell,m}(\hat{\bm{k}})\sum_{s} \frac{\rho(M_2)}{(2s)!} e^{-||\alpha||^2} ||\alpha||^{4s} |g^{(\ell)}|^2\tilde{\mathcal{P}}^{(\ell)}(\hat{\bm{k}}', \hat{\bm{k}})
\nn
&=\frac{|\omega|}{8\pi M_2} \sum_s \frac{\rho(M_2)}{(2s)!} e^{-||\alpha||^2} ||\alpha||^{4s} |g^{(\ell)}|^2
\nn
&=\frac{|\omega|}{8\pi M_2} \rho(M_2) |g^{(\ell)}|^2
\,,
\end{align}
and the emission probability is
\begin{align}
    P^{\rm em}_{\ell,m} &=\frac{|\omega|}{8\pi M_1} \rho(M_1) |g^{(\ell)}|^2
    \,,
\end{align}
where the remaining asymmetry between the two comes from the difference in the phase space volume of the final states.

\section{Greybody factors and relation with emission and absorption probabilities}
\label{sec:greybody}
One of the key equations in our derivation of the thermal spectrum was the use of \eqref{balance_eq}, connecting the classical transmissivity, also known as the greybody factor, with the quantum absorption and emission probabilities. In this section, we provide a derivation of this equation by studying a classical wave scattering based on quantum scattering amplitudes. Note that the same equation is discussed in Ref.~\cite{Endlich:2016jgc} under the assumption of weak coupling. The purpose of this section is to obtain \eqref{balance_eq} without using an expansion of couplings because we are interested in objects with strong absorption.

We start by considering an isolated object described as in Ref.~\cite{Cristofoli:2021vyo} by a localized massive wavepacket and by an incoming wave made, for simplicity, of massless scalars 
\begin{equation}
    \ket{\Psi;\beta}:=e^{-\frac{1}{2}\| \beta \|^2}  \sum_{n=0}^{+\infty} \frac{1}{n!} \int d\Phi(p,k_1,...k_{n}) \beta(k_1)...\beta(k_n)\phi(p) \ket{p,\alpha;k_1,...,k_n}
\end{equation}
with
\begin{align}
    \| \beta \|^2 := \int d\Phi(k)|\beta(k)|^2
    \,.
\end{align}
Here, $\beta(k)$ is a coherent wave shape of the incoming wave, whose magnitude can be read by computing the expectation value of the number operator
\begin{align}
    \bra{\Psi;\beta}\hat{N}\ket{\Psi;\beta} = \|\beta\|^2\,, \qquad 
    \hat{N}:=\int \dd \Phi(k) \hat{a}^{\dagger}_k \hat{a}_k
    \,.
\end{align}
Hence, the intensity change of the coherent wave after scattering is given by
\begin{align}
\bra{\Psi;\beta}\hat{S}^{\dagger}\hat{N}\hat{S}\ket{\Psi;\beta}-{\bra{\Psi;\beta}\hat{N}\ket{\Psi;\beta}} = - \int \dd \Phi(k) |\mathcal{T}(k)|^2 |\beta(k)|^2
    \,,
\end{align}
where $|\mathcal{T}(k)|>0$ represents a classical absorption rate of the coherent wave at the momentum $k$. Using unitarity, the left-hand side can be expressed as
\begin{align}
    \bra{\Psi;\beta}\hat{S}^{\dagger}\hat{N}\hat{S}\ket{\Psi;\beta}-\bra{\Psi;\beta}\hat{N}\ket{\Psi;\beta} = \bra{\Psi;\beta}\hat{S}^{\dagger}[\hat{N},\hat{S}]\ket{\Psi;\beta}
    \label{number_change}
\end{align}
which we would like to compute in the following.

We begin by considering the commutator $[\hat{N},\hat{S}]$ sandwiched by $n$-particle states:
\begin{align}
    \bra{p',s';k'_1;\cdots; k_m'}[\hat{N},\hat{S}]\ket{p,s;k_1;\cdots ;k_n}=(m-n)\bra{p',s';k'_1;\cdots; k_m'}\hat{S}\ket{p,s;k_1;\cdots ;k_n}
\end{align}
which manifests that only the number change processes contribute to this quantity. Similarly, using the completeness relation, we find
\begin{align}
    &\Delta N_{n',n}
    \nn
    &=\bra{p',s';k'_1;\cdots; k'_{n'}}\hat{S}^{\dagger}
    [\hat{N},\hat{S}]\ket{p,s;k_1;\cdots ;k_n}
    \nn
    &=\sum_m \frac{1}{m!} \sum_{s''} \int \dd \mu^2 \int \dd\Phi(p'',k''_1,\cdots, k''_m)\rho(\mu,s'')
    \nn
    &\qquad \qquad \times \bra{p',s';k'_1;\cdots; k'_{n'}}\hat{S}^{\dagger}\ket{p'',s'';k''_1;\cdots;k''_m}\bra{p'',s'';k''_1;\cdots;k''_m}[\hat{N},\hat{S}]\ket{p,s;k_1;\cdots; k_n}
    \nn
    &=i\hdelta^{(4)}(P_n-P'_n)\delta_{n',n}\left(-\sum
    \begin{tikzpicture}[baseline=-2]
\begin{feynhand}
\draw [
    thick,
    decoration={
        brace,
    },
    decorate
] (1.6, 0.6) -- (1.6,-0.25) ;
\node at (1.9, 0.175) {$n$};
\foreach \y in {0,0.55}
\propag[boson] (-1.5, \y)--(1.5,\y);
\node at (0, 0.35) {$\vdots$};
\draw (-1.5,-0.6) -- (-0.6,-0.6);
\draw (1.5,-0.6) -- (0.6,-0.6);
\draw[very thick] (-0.6,-0.4)--(0.6,-0.4);
\propag[boson] (-1.5, -0.2) -- (-0.6, -0.2);
\bub{-0.6}{-0.4}{0.3}{-}{}
\propag[boson] (1.5, -0.2) -- (0.6, -0.2);
\bub{0.6}{-0.4}{0.3}{+}{}
\end{feynhand}
\end{tikzpicture}
+\sum
 \begin{tikzpicture}[baseline=-2]
\begin{feynhand}
\draw [
    thick,
    decoration={
        brace,
    },
    decorate
] (1.6, 0.6) -- (1.6,-0.25) ;
\node at (1.9, 0.175) {$n$};
\foreach \y in {0,0.55}
\propag[boson] (-1.5, \y)--(1.5,\y);
\node at (0, 0.35) {$\vdots$};
\draw (-1.5,-0.6) -- (-0.6,-0.6);
\draw (1.5,-0.6) -- (0.6,-0.6);
\draw[very thick] (-0.6,-0.7)--(0.6,-0.7);
\propag[boson] (-1.5, -0.2) -- (0.5, -0.45);
\propag[boson] (1.5, -0.2) -- (-0.5, -0.45);
\bub{-0.6}{-0.65}{0.3}{-}{}
\bub{0.6}{-0.65}{0.3}{+}{}
\end{feynhand}
\end{tikzpicture}
+\sum
 \begin{tikzpicture}[baseline=-2]
\begin{feynhand}
\draw [
    thick,
    decoration={
        brace,
    },
    decorate
] (1.6, 0.6) -- (1.6,-0.25) ;
\node at (1.9, 0.175) {$n$};
\foreach \y in {-0.2, 0,0.55}
\propag[boson] (-1.5, \y)--(1.5,\y);
\node at (0, 0.35) {$\vdots$};
\draw (-1.5,-0.6) -- (-0.6,-0.6);
\draw (1.5,-0.6) -- (0.6,-0.6);
\draw[very thick] (-0.6,-0.8)--(0.6,-0.8);
\propag[boson] (-0.6, -0.4) -- (0.6, -0.4);
\bub{-0.6}{-0.6}{0.3}{-}{}
\bub{0.6}{-0.6}{0.3}{+}{}
\end{feynhand}
\end{tikzpicture}
\right)
\nn
&+\cdots\,,
\label{DeltaN}
\end{align}
where $P_n=p+\sum^n_{i=1} k$, $P'=p'+\sum_{i=1}^n k'$, and the ellipsis stands for contributions from higher-point amplitudes. Here, we use the similar diagrammatic notation used in Chapter 4 of Ref.~\cite{Eden:1966dnq}: the $\pm$ bubbles denote the $(-1)\times(\text{scattering amplitudes})$ and their conjugate, all lines represent on-shell particles with the solid and wavy lines being massive and massless particles, the bold line is the mass-changed state, and the summation is taken over all possible choices of different particles with the same topology of the diagram. The first term describes absorptions; the second term can be understood as induced emissions; and the third one is spontaneous emissions. Crucially, the first two terms have $n\times n!$ different pairings while the third one has only $n!$ pairings. Hence, the third term can be negligible in the large $n$ limit, corresponding to the coherent wave with a classical intensity. As a result, we have
\begin{align}
    \Delta N_{n',n}&\simeq \hdelta^{(4)}(P_n-P_n') \delta_{n',n}\sum_{s''}\int \dd^2 \mu \int \dd\Phi(p'') \rho(\mu,s'')
    \nn
    &\times \sum_{i'=1}^n \sum_{i=1}^n \left(-\Amp^*_{p'' \leftarrow p',k'_{i'} } \Amp_{p'' \leftarrow p, k_i} +\Amp^*_{p'',k_i \leftarrow p' } \Amp_{p'',k'_{i'} \leftarrow p} 
    \right)
    \nn
    &\times \sum_{\sigma } \{\hdelta_{\Phi} (k'_1-k_{\sigma_1})\cdots \hdelta_{\Phi}(k'_{i-1}-k_{\sigma_{i'-1}})\hdelta_{\Phi}(k'_{i+1}-k_{\sigma_{i'+1}})\cdots \hdelta_{\Phi}(k'_n-k_{\sigma_n})\}
    \nn
    &+\cdots,
\end{align}
where the summation of $\sigma$ is over $(n-1)!$ permutations, and $\hdelta_{\Phi}(k-k')=2\omega \hdelta^{(3)}(\bm{k}-\bm{k}')$ is the on-shell delta function. Then, the intensity change \eqref{number_change} is given by
\begin{align}
    &\bra{\Psi;\beta}\hat{S}^{\dagger}[\hat{N},\hat{S}]\ket{\Psi;\beta}
    \nn
    &\simeq \int \dd \Phi(p,p', k, k') \phi^*(p')\phi(p) e^{-\|\beta|^2} \sum_{n} \frac{\|\beta \|^{2n-2}}{(n-1)!}
    \beta^*(k')\beta(k) \left(-\frac{\Disc_s\Amp_{1'\leftarrow 1}}{2\pi i}+ \frac{\Disc_u \Amp_{1'\leftarrow 1}}{2\pi i} \right) \hdelta^{(4)}(p+k-p'-k')+\cdots
    \nn
    &=\int \dd \Phi(p,p', k, k') \phi^*(p')\phi(p) 
    \beta^*(k')\beta(k) \left(-\frac{\Disc_s\Amp_{1'\leftarrow 1}}{2\pi i}+ \frac{\Disc_u \Amp_{1'\leftarrow 1}}{2\pi i} \right)\hdelta^{(4)}(p+k-p'-k') +\cdots
\end{align}
where $\Disc_s$ and $\Disc_u$ denote the discontinuities associated with the $s$-channel cut (the fist term in \eqref{DeltaN}) and the $u$-channel cut (the second term), respectively.

We choose the initial coherent wave as a partial wave of $\ell$:
\begin{align}
    \beta(k)=\frac{\sqrt{2\omega}}{4\pi}\gamma(\omega)Y_{\ell,m}(\hat{\bm{k}})
    \,, \qquad \|\beta \|^2 = \int \hd \omega\, |\gamma(\omega)|^2
    \,.
\end{align}
We also assume that classical physics is insensitive to an $\mathcal{O}(\hbar)$ change of the initial mass of the macroscopic object
\begin{align}
    \Disc_u \Amp_{1'\leftarrow 1} \simeq \Disc_u \Amp_{2'\leftarrow 2}
    \,.
\end{align}
Then, by taking the classical limit, the intensity change \eqref{number_change} is
\begin{align}
    \bra{\Psi;\beta}\hat{S}^{\dagger}[\hat{N},\hat{S}]\ket{\Psi;\beta}= \int \hd \omega |\gamma(\omega)|^2(-P^{\rm abs}_{\ell,m}(\omega)+P^{\rm em}_{\ell,m}(-\omega)) + \cdots\,,
\end{align}
yielding
\begin{align}
    P^{\rm abs}_{\ell,m}(\omega)-P^{\rm em}_{\ell,m}(-\omega)+\cdots = |\mathcal{T}_{\ell}(\omega)|^2
    \,,
\end{align}
where $|\mathcal{T}_{\ell}|$ is the transmissivity of the partial wave $\ell$.

Now we discuss when the contributions from higher-point amplitudes may be ignored. We stress that we cannot use the expansion in terms of coupling constants, especially for absorption, because we are interested in situations with strong absorption $P^{\rm abs}_{\ell,m}\sim \rho(M_2)|g|^2 = \mathcal{O}(1)$. As an example, we consider a two-particle absorption contribution, which is diagrammatically represented by
\begin{align}
        \begin{tikzpicture}[baseline=-2]
\begin{feynhand}
\draw [
    thick,
    decoration={
        brace,
    },
    decorate
] (1.6, 0.6) -- (1.6,-0.45) ;
\node at (1.9, 0.175) {$n$};
\foreach \y in {0,0.55}
\propag[boson] (-1.5, \y)--(1.5,\y);
\node at (0, 0.35) {$\vdots$};
\draw (-1.5,-0.8) -- (-0.6,-0.8);
\draw (1.5,-0.8) -- (0.6,-0.8);
\draw[very thick] (-0.6,-0.5)--(0.6,-0.5);
\propag[boson] (-1.5,-0.4) -- (-0.4, -0.4);
\propag[boson] (-1.5, -0.2) -- (-0.6, -0.2);
\bub{-0.6}{-0.5}{0.4}{-}{}
\propag[boson] (1.5,-0.4) -- (0.4, -0.4);
\propag[boson] (1.5, -0.2) -- (0.6, -0.2);
\bub{0.6}{-0.5}{0.4}{+}{}
\end{feynhand}
\end{tikzpicture}
\,.
\end{align}
There are $\frac{n(n-1)}{4}\times n!$ diagrams of this topology. The correction to the intensity change is
\begin{align}
    &-\int \dd \Phi(p,p',k_1,k_2,k_1',k_2') \phi^*(p')\phi(p) e^{-\|\beta |^2} \sum_n \frac{1}{2}\frac{\|\beta\|^{2n-4}}{(n-2)!}\beta^*(k'_1)\beta^*(k'_2) \beta(k_1)\beta(k_2) \frac{\Disc_s \Amp_6}{2\pi i} \hdelta^{(4)}(P_2-P'_2)
    \nn
    &=-\int \dd \Phi(p,p',k_1,k_2,k_1',k_2') \phi^*(p')\phi(p) \frac{1}{2}\beta^*(k'_1)\beta^*(k'_2) \beta(k_1)\beta(k_2) \frac{\Disc_s \Amp_6}{2\pi i} \hdelta^{(4)}(P_2-P'_2)
\end{align}
where $\Disc_s \Amp_6$ is the discontinuity of the six-point amplitude. This contribution is of the order of $\|\beta \|^4$ and represents a nonlinear correction due to a large intensity of the coherent wave. Therefore, we can ignore the contributions from higher-point absorption amplitudes when the initial intensity of the classical wave is sufficiently small, even if the object exhibits a strong absorption. Note that a similar conclusion does not hold for strong emission. For example, the number of the following mixed induced and spontaneous emission diagrams
\begin{align}
    \begin{tikzpicture}[baseline=-2]
\begin{feynhand}
\draw [
    thick,
    decoration={
        brace,
    },
    decorate
] (1.6, 0.6) -- (1.6,-0.25) ;
\node at (1.9, 0.175) {$n$};
\foreach \y in {0,0.55}
\propag[boson] (-1.5, \y)--(1.5,\y);
\node at (0, 0.35) {$\vdots$};
\draw (-1.5,-0.75) -- (-0.6,-0.75);
\draw (1.5,-0.75) -- (0.6,-0.75);
\draw[very thick] (-0.6,-1)--(0.6,-1);
\propag[boson] (-0.6,-0.75) -- (0.6,-0.75);
\propag[boson] (-1.5, -0.2) -- (0.5, -0.5);
\propag[boson] (1.5, -0.2) -- (-0.5, -0.5);
\bub{-0.6}{-0.75}{0.35}{-}{}
\bub{0.6}{-0.75}{0.35}{+}{}
\end{feynhand}
\end{tikzpicture}
\end{align}
is $n\times n!$, which will thus be a correction at $\mathcal{O}(\|\beta \|^2)$ for coherent waves. Hence, neglecting higher-point emission amplitudes requires assuming that emissions are sufficiently weak. All in all, \eqref{balance_eq} holds under the conditions that (i) the initial intensity of the classical wave is small and (ii) emission processes are weak, but it does not require the assumption of weak absorption. This conclusion is consistent with the observation in Ref.~\cite{Aoki:2025ihc} that \eqref{balance_eq} holds well approximately even for a strong absorption regime, corresponding to frequencies $\omega \gtrsim 1/GM$, within the linear perturbation theory around the Schwarzschild BH in both Boulware and Unruh vacuum.

\end{widetext}
%----------------------------------------------------------------------------

\end{document}